\documentclass[letterpaper,english,preprint]{aastex}
\usepackage[T1]{fontenc}
\usepackage[latin1]{inputenc}

\makeatletter
\AtBeginDocument{
  
}

\usepackage{babel}
\makeatother
\begin{document}

\title{Polarizing grids, their assemblies and beams of radiation.}

\author{Martin Houde\altaffilmark{1,2},\email{houde@ulu.submm.caltech.edu}
Rachel L. Akeson\altaffilmark{3}, John E. Carlstrom\altaffilmark{4},
James W. Lamb\altaffilmark{5}, David A. Schleuning\altaffilmark{4},
David P. Woody\altaffilmark{5}}

\altaffiltext{1}{Caltech Submillimeter Observatory, 111 Nowelo Street, Hilo, HI 96720}

\altaffiltext{2}{Département de Physique, Université de Montréal, Montréal, Québec H3C 3J7, Canada}

\altaffiltext{3}{Infrared Processing \& Analysis Center, California Institute of Technology, Pasadena, CA 91125}

\altaffiltext{4}{Department of Astronomy and Astrophysics and Enrico Fermi Institute, University of Chicago, Chicago, IL 60637}

\altaffiltext{5}{Owens Valley Radio Observatory, California Institute of Technology, Big Pine, CA 93513}

\begin{abstract}
This article gives an analysis of the behavior of polarizing grids
and reflecting polarizers by solving Maxwell's equations, for arbitrary
angles of incidence and grid rotation, for cases where the excitation
is provided by an incident plane wave or a beam of radiation. The
scattering and impedance matrix representations are derived and used
to solve more complicated configurations of grid assemblies. The results
are also compared with data obtained in the calibration of reflecting
polarizers at the Owens Valley Radio Observatory (OVRO). From these
analysis, we propose a method for choosing the optimum grid parameters
(wire radius and spacing). We also provide a study of the effects
of two types of errors (in wire separation and radius size) that can
be introduced in the fabrication of a grid. 
\end{abstract}


\section{Introduction.}

The literature on wire grids is abundant and they have been studied
with different techniques and for numerous applications. Most of the
analysis were however restricted to special cases of incident field
and grid orientations. The more general and arbitrary situation seems
to have been first studied by Wait (see \citet{Wait 1955a} and \citet{Larsen}).
This problem is addressed again in this paper and follows a line of
analysis fairly similar to the one used by Wait. Our treatment is,
however, more general in that we do not assume that the wires of the
grid are induced with only a longitudinal current; we will indeed
show that an azimuthal component is also present. We also solve for
the induced current by considering the tangential components of both
the electric and magnetic fields at the surface of the wires.

This analysis is carried out in the next two sections and will serve
as our basis for the treatment of the reflecting polarizer (section
\ref{sec:RP}) and the introduction of the scattering and impedance
matrix representations for a grid (section \ref{sec:imp}) which will
in turn enable us to briefly discuss more complicated systems. These
matrices will be particularly useful in allowing us to define what
will be called the principal axes of a grid. These are two orthogonal
and independent directions of polarization in the plane of the incident
radiation along which an arbitrary electric field can be decomposed
and shown to scatter without cross-polarization. With this representation
at hand, it will then be possible to derive a set of optimal parameters
(wire radius and spacing) to be used in the selection of a grid. We
will also present an analysis of the effects of random errors that
can be introduced in the fabrication of grids, the results obtained
will then be compared to experimental results previously published
by \citet{Shapiro}.

The last section will be dedicated to the study of the more subtle
impacts that the nature of the incoming radiation can have on the
response of a grid assembly such as a reflecting polarizer (section
\ref{sec:RP}). Although limited to this particular case, our discussion
could possibly apply to other types of instruments. We have also included
at the end (Appendix B) a list of the symbols used in the different
equations.

\section{The case of a single wire.\label{sec:wire}}

Before trying to solve the problem of the grid or the reflecting polarizer,
it is preferable to study the case of a single conducting wire. It
will serve as the basis for our studies of the more complicated cases
to follow in subsequent sections.

Let's suppose that a wire of radius $a$ is oriented, as depicted
in Figure \ref{fig:wire}, parallel to the $x$-axis at $y=y_{o}$,
$z=z_{o}$ and that it is subjected to an incident plane wave $\mathbf{E}_{i}(\mathbf{r})$
of arbitrary direction and polarization:

\begin{equation}
\mathbf{E}_{i}(\mathbf{r})=E_{o}(\alpha'\mathbf{e}_{x}+\beta'\mathbf{e}_{y}+\gamma'\mathbf{e}_{z})\exp(-j(\mathbf{k}\cdot\mathbf{r}-\omega t))\label{eq:inc}\end{equation}

\noindent with 

\[
\mathbf{k}=k(\alpha\mathbf{e}_{x}+\beta\mathbf{e}_{y}+\gamma\mathbf{e}_{z})\]

\noindent and where, of course, the following conditions of normalization
and orthogonality apply: $\alpha^{2}+\beta^{2}+\gamma^{2}=\alpha'^{2}+\beta'^{2}+\gamma'^{2}=1$
and $\alpha\alpha'+\beta\beta'+\gamma\gamma'=0$. Using the coordinate
system depicted in Figure \ref{fig:wire} we have $\alpha=\sin(\chi_{i})\sin(\varphi_{g})$,
$\beta=\sin(\chi_{i})\cos(\varphi_{g})$ and $\gamma=\cos(\chi_{i})$
where $\chi_{i}$ is the angle of incidence and $\varphi_{g}$ the
angle of grid rotation.

In everything that follows, we will drop the $\exp(j\omega t)$ term
and assume it to be implicit in the equations. We will also suppose
that the wire is of infinite length and made of a good conducting
material of conductivity $\sigma$ such that any current flowing through
it can be accurately represented by a surface current vector $\mathbf{K}$.
This quantity is related to the current density $\mathbf{J}(\mathbf{r})$
as follows:

\begin{equation}
\mathbf{J}(\mathbf{r})=\mathbf{K}\,\delta(\rho-a)\,\exp(-j\mathbf{k}\cdot\mathbf{r})\label{eq:cur}\end{equation}

\noindent where

\[
\mathbf{K}=K^{x}\mathbf{e}_{x}+K^{\theta}\mathbf{e}_{\theta}\]

\noindent and $y-y_{o}=\rho\cos(\theta)$, $z-z_{o}=\rho\sin(\theta)$.

Before we solve for the scattered fields, it is to our advantage to
note that for the case considered here (i.e., thin wire with an approximate
solution involving no angular mode dependency), the problem can be
broken in two parts or modes. The mode where the electrical field
is parallel to the plane defined by $\mathbf{e}_{x}$ and $\mathbf{k}$
(the transverse magnetic or TM-mode) is related to the presence of
$K^{x}$ while another, where the magnetic field is parallel to this
same plane, the transverse electric or TE-mode is related to $K^{\theta}$.
The analysis will, therefore, be facilitated with the use of the two
vector potentials $\mathbf{A}_{s}$ and $\mathbf{F}_{s}$ for the
scattered fields \citep{Balanis}. 

The TM-mode can be analyzed using the vector potential $\mathbf{A}_{s}$,
in the Lorentz gauge, with $\mathbf{F}_{s}=0$. The needed equations
are:

\begin{eqnarray}
\mathbf{A}_{s}(\mathbf{r}) & = & \frac{\mu_{o}}{4\pi}\int\mathbf{J}(\mathbf{r}')\,\frac{\exp(-jkR)}{R}d^{3}r'\label{eq:A}\\
\mathbf{E}_{s}(\mathbf{r}) & = & \frac{c^{2}}{j\omega}\nabla(\nabla\cdot\mathbf{A}_{s}(\mathbf{r}))-j\omega\mathbf{A}_{s}(\mathbf{r})\label{eq:Esw}\\
\mathbf{H}_{s}(\mathbf{r}) & = & \frac{1}{\mu_{o}}\nabla\times\mathbf{A}_{s}(\mathbf{r})\label{eq:Hsw}\end{eqnarray}

\noindent with \[
R^{2}=(x-x')^{2}+(y-y')^{2}+(z-z')^{2}\,.\]

Since we are concerned here with the longitudinal component of the
surface current density, we only need to consider the $A^{x}$ component
of the vector potential (i.e., we set $A^{\rho}=A^{\theta}=0$). Equation
(\ref{eq:A}) can be solved exactly when $\mathbf{K}$ is expanded
with a Fourier series, but in cases where the wavelength of the incident
wave is much larger than the wire radius it can be shown that: 

\begin{equation}
A_{s}^{x}(\mathbf{r})=\frac{\pi\mu_{o}a}{2j}K^{x}\, H_{0}^{(2)}(k'\rho)\exp(-j\varphi)\label{eq:Asol}\end{equation}

\noindent with $k'=k\sqrt{1-\alpha^{2}}$, $\varphi=k(\alpha x+\beta y_{o}+\gamma z_{o})$
and where $H_{n}^{(2)}(x)$ is Hankel's function of the second kind
of order $n$.

On the other hand, it is advantageous to study the TE-mode with the
vector potential \textbf{$\mathbf{F}_{s}$,} in the appropriate gauge,
with $\mathbf{A}_{s}=0$ \citep{Balanis}. To do so, we will not consider
the effect of the current density (more precisely its azimuthal component)
but that of the magnetization vector $\mathbf{M}$ that it induces.
The relevant equations are now:

\begin{eqnarray}
\mathbf{J}(\mathbf{r}) & = & \nabla\times\mathbf{M}(\mathbf{r})\label{eq:JM}\\
\mathbf{F}_{s}(\mathbf{r}) & = & \frac{j\omega\mu_{o}\varepsilon_{o}}{4\pi}\int\mathbf{M}(\mathbf{r}')\,\frac{\exp(-jkR)}{R}d^{3}r'\label{eq:Fs}\\
\mathbf{E}_{s}(\mathbf{r}) & = & -\frac{1}{\varepsilon_{o}}\nabla\times\mathbf{F}_{s}(\mathbf{r})\label{eq:EsF}\\
\mathbf{H}_{s}(\mathbf{r}) & = & \frac{c^{2}}{j\omega}\nabla(\nabla\cdot\mathbf{F}_{s}(\mathbf{r}))-j\omega\mathbf{F}_{s}(\mathbf{r})\label{eq:HsF}\end{eqnarray}

\noindent with $R$ as defined above.

Since we are now concerned with the azimuthal component of the surface
current density, we only need to consider the $M^{x}$, $F_{s}^{x}$
components of the magnetization and vector potential (i.e., we have
$M^{\rho}=M^{\theta}=F_{s}^{\rho}=F_{s}^{\theta}=0$). Again, in cases
where the wavelength of the incident wave is much larger than the
wire radius it can be shown that: 

\begin{equation}
F_{s}^{x}(\mathbf{r})=\frac{\pi\omega\mu_{o}\varepsilon_{o}a^{2}}{4}K^{\theta}\, H_{0}^{(2)}(k'\rho)\exp(-j\varphi)\,.\label{eq:Fsol}\end{equation}

It is now straightforward to calculate the scattered fields by combining
the solution obtained for each mode (using equations (\ref{eq:Esw}),
(\ref{eq:Hsw}) and (\ref{eq:Asol}) for the TM-mode and equations
(\ref{eq:EsF}), (\ref{eq:HsF}) and (\ref{eq:Fsol}) for the TE-mode):

\begin{eqnarray}
E_{s}^{\rho}(\mathbf{r}) & = & -j\alpha\sqrt{1-\alpha^{2}}F\, K^{x}H_{1}^{(2)}(k'\rho)\exp(-j\varphi)\label{eq:Er}\\
E_{s}^{\theta}(\mathbf{r}) & = & -\sqrt{1-\alpha^{2}}F\,\frac{ka}{2}\, K^{\theta}H_{1}^{(2)}(k'\rho)\exp(-j\varphi)\label{eq:Et}\\
E_{s}^{x}(\mathbf{r}) & = & -\left(1-\alpha^{2}\right)F\, K^{x}H_{0}^{(2)}(k'\rho)\exp(-j\varphi)\label{eq:Ex}\\
H_{s}^{\rho}(\mathbf{r}) & = & \frac{\alpha\sqrt{1-\alpha^{2}}}{Z_{o}}F\,\frac{ka}{2}\, K^{\theta}H_{1}^{(2)}(k'\rho)\exp(-j\varphi)\label{eq:Hr}\\
H_{s}^{\theta}(\mathbf{r}) & = & -j\frac{\sqrt{1-\alpha^{2}}}{Z_{o}}F\, K^{x}H_{1}^{(2)}(k'\rho)\exp(-j\varphi)\label{eq:Ht}\\
H_{s}^{x}(\mathbf{r}) & = & -j\frac{\left(1-\alpha^{2}\right)}{Z_{o}}F\,\frac{ka}{2}\, K^{\theta}H_{0}^{(2)}(k'\rho)\exp(-j\varphi)\label{eq:Hx}\end{eqnarray}

\noindent where $F=\frac{\pi\mu_{o}\omega a}{2}$ and $Z_{o}=\sqrt{\frac{\mu_{o}}{\varepsilon_{o}}}$
is the impedance of free space. Note that although equations (\ref{eq:Er})-(\ref{eq:Hx})
represent the scattered field, the components of surface current density
that are included in these equations are that of the, yet undetermined,
total surface current density which we are now in a position to evaluate. 

In order to do so, we must first express the incident plane wave in
the appropriate coordinate system. This can be done by first using
the following expression:

\begin{equation}
\exp(-jk(\beta y+\gamma z))=\sum_{n=-\infty}^{\infty}(-j)^{n}J_{n}(k'\rho)\exp(jn\theta')\label{eq:bes}\end{equation}

\noindent with $\theta'=\theta-\arctan\left(\frac{\gamma}{\beta}\right)$,
$J_{n}\left(x\right)$ the Bessel function of order $n$ and by again
splitting the incident field in the two modes defined earlier \citep{van de Hulst, Balanis}.
This enables us to express the plane wave in cylindrical coordinates
and match the fields with the usual boundary conditions for their
tangential components at the surface of the wire. For the TM-mode
the condition is:

\begin{equation}
E_{i}^{x}+E_{s}^{x}=Z_{s}(H_{i}^{\theta}+H_{s}^{\theta})\,.\label{eq:bcx}\end{equation}

Within the order of precision used for our analysis ($\lambda\gg a$)
and considering a solution with no angular dependency, it can be shown
that: 

\begin{eqnarray*}
E_{i}^{x} & \simeq & \alpha'E_{o}\exp\left(-jk\left(\alpha x+\gamma z_{o}\right)\right)\\
H_{i}^{\theta} & \simeq & j\alpha'\frac{E_{o}}{Z_{o}}\cdot\frac{ka}{2}\exp\left(-jk\left(\alpha x+\gamma z_{o}\right)\right)\,.\end{eqnarray*}

For the TE-mode we have:

\begin{equation}
E_{i}^{\theta}+E_{s}^{\theta}=-Z_{s}(H_{i}^{x}+H_{s}^{x})\label{eq:bct}\end{equation}

\noindent with:

\begin{eqnarray*}
E_{i}^{\theta} & \simeq & -j\left(\gamma'\beta-\beta'\gamma\right)E_{o}\frac{ka}{2}\exp\left(-jk\left(\alpha x+\gamma z_{o}\right)\right)\\
H_{i}^{x} & \simeq & \left(\gamma'\beta-\beta'\gamma\right)\frac{E_{o}}{Z_{o}}\exp\left(-jk\left(\alpha x+\gamma z_{o}\right)\right)\,.\end{eqnarray*}

In equations (\ref{eq:bcx}) and (\ref{eq:bct}), $Z_{s}=(1+j)\sqrt{\frac{\mu_{o}\omega}{2\sigma}}$
is the surface impedance of the wire \citep{Jackson}. It is to be
noted that for wires of small radius, relative to the wavelength,
the boundary conditions (\ref{eq:bcx}) and (\ref{eq:bct}) along
with the equation for $Z_{s}$ represent approximations that are only
valid in the lowest mode and for a sufficiently good conductor. A
more rigorous treatment shows that these equations will be modified
in the more general case \citep{Wait 1979, Bouche}. But for the purpose
of our analysis, the approximation used here is adequate.

When solving these two sets of equations we find the following expressions
for the components of the total surface current densities:

\begin{eqnarray}
K^{x} & = & \frac{E_{o}}{F}\cdot\frac{\alpha'\left(1-j\frac{Z_{s}}{Z_{o}}\cdot\frac{ka}{2}\right)}{\left(1-\alpha^{2}\right)H_{0}^{(2)}(k'a)-j\frac{Z_{s}}{Z_{o}}\sqrt{1-\alpha^{2}}H_{1}^{(2)}(k'a)}\label{eq:Kxw}\\
K^{\theta} & = & \frac{E_{o}}{F}\cdot\frac{-j(\gamma'\beta-\beta'\gamma)\left(1+j\frac{Z_{s}}{Z_{o}}\cdot\frac{2}{ka}\right)}{\sqrt{1-\alpha^{2}}H_{1}^{(2)}(k'a)+j\frac{Z_{s}}{Z_{o}}\left(1-\alpha^{2}\right)H_{0}^{(2)}(k'\rho)}\,.\label{eq:Ktw}\end{eqnarray}

These last two equations can be inserted in equations (\ref{eq:Er})-(\ref{eq:Hx})
to calculate the value of the fields at any point exterior to the
wire. For a good conductor the internal fields are practically nonexistent.
Equations (\ref{eq:Kxw}) and (\ref{eq:Ktw}) are in agreement with
the results presented in \citet[see ch. 11]{Balanis} for the case
of normal incidence and a perfectly conducting wire.

\section{The polarizing grid.}

\subsection{Analysis.\label{sec:grid}}

With the solution for a single wire in hand, the problem of a configuration
of an infinite number of wires of infinite length separated by a distance
$d$ is simplified if one realizes that every wire will be induced
with the same surface current $\mathbf{K}$. The only difference will
be a phase term in the current density $\mathbf{J}(\mathbf{r})$,
given by equation (\ref{eq:cur}), which depends on the position of
the wire along the $y$-axis. The same thing can be said for the scattered
fields from any given wire, one only has to replace $y_{o}$ by $nd$
in equations (\ref{eq:Er})-(\ref{eq:Hx}), where $n$ is an integer
that determines the position of the wire.

If the scattered fields are now just the sum of all the different
scattered fields from the individual wires, care must however be taken
in evaluating the surface current. First, when one matches the boundary
conditions it must be done simultaneously at the surface of every
wire. However, since we are dealing with an infinite number of infinitely
long wires subjected to the same incident plane wave, it turns out
that it is sufficient to do so for only one of the wires. If the boundary
conditions are matched for one wire they will be for all. We have
chosen for our calculations the {}``center'' wire at $n=0$. Second,
to match the boundary conditions we must express the scattered fields
of each and every wire in a cylindrical coordinate system centered
on the position of this {}``center'' wire.

When this is done, we find the following expressions for the components
of the induced total surface current density:

\begin{eqnarray}
K^{x} & = & \frac{E_{o}}{F}\cdot\alpha'\frac{N_{x}}{\Delta_{x}}\label{eq:Kxg}\\
K^{\theta} & = & -j\frac{E_{o}}{F}\cdot(\gamma'\beta-\beta'\gamma)\frac{N_{\theta}}{\Delta_{\theta}}\label{eq:Ktg}\end{eqnarray}

\noindent with 

\begin{eqnarray}
N_{x} & = & 1-j\frac{Z_{s}}{Z_{o}}\cdot\frac{ka}{2}\label{eq:Nx}\\
\Delta_{x} & = & \left(1-\alpha^{2}\right)S_{1}-j\frac{Z_{s}}{Z_{o}}\cdot\sqrt{1-\alpha^{2}}H_{1}^{(2)}(k'a)\label{eq:deltax}\\
N_{\theta} & = & 1+j\frac{Z_{s}}{Z_{o}}\cdot\frac{2}{ka}\label{eq:Nt}\\
\Delta_{\theta} & = & \sqrt{1-\alpha^{2}}H_{1}^{(2)}(k'a)+j\frac{Z_{s}}{Z_{o}}\cdot\left(1-\alpha^{2}\right)S_{1}\label{eq:deltat}\end{eqnarray}

\noindent and

\begin{equation}
S_{1}=H_{0}^{(2)}(k'a)+2\sum_{n=1}^{\infty}H_{0}^{(2)}(k'nd)\cos(k\beta nd)\,.\label{eq:S1}\end{equation}

We will give in section \ref{sec:approx} adequate approximations
for $\Delta_{x}$ and $\Delta_{\theta}$ that will greatly simplify
the evaluation of the reflection and transmission coefficients which
are soon to follow.

By using the appropriate expansions for series of Hankel's functions
we can write down the expressions for the components of the total
electric field far away from the grid:

\begin{eqnarray*}
E_{T}^{x}(\mathbf{r}) & = & \alpha'E_{o}\exp(-j\mathbf{k}\cdot\mathbf{r})-\frac{\left(1-\alpha^{2}\right)}{\gamma}\cdot\frac{\lambda F}{\pi d}K^{x}\exp(-jk\gamma|z-z_{o}|)\exp(-j\varphi)\\
E_{T}^{y}(\mathbf{r}) & = & \beta'E_{o}\exp(-j\mathbf{k}\cdot\mathbf{r})\\
 &  & +\frac{\lambda F}{\pi d}\left[\frac{\alpha\beta}{\gamma}K^{x}+j\frac{ka}{2}\, K^{\theta}\frac{z-z_{o}}{|z-z_{o}|}\right]\exp(-jk\gamma|z-z_{o}|)\exp(-j\varphi)\\
E_{T}^{z}(\mathbf{r}) & = & \gamma'E_{o}\exp(-j\mathbf{k}\cdot\mathbf{r})\\
 &  & +\frac{\lambda F}{\pi d}\left[\alpha K^{x}\frac{z-z_{o}}{|z-z_{o}|}-j\frac{\beta}{\gamma}\cdot\frac{ka}{2}\, K^{\theta}\right]\exp(-jk\gamma|z-z_{o}|)\exp(-j\varphi)\end{eqnarray*}

\noindent where $\varphi=k(\alpha x+\beta y+\gamma z_{o})$. From
these it is now straightforward to get the reflection and transmission
coefficients (normalized to $E_{o}$) in the far-field:

\begin{eqnarray}
R^{x} & = & -\frac{F}{E_{o}}\cdot\frac{\lambda}{\pi d}\cdot\frac{\left(1-\alpha^{2}\right)}{\gamma}\, K^{x}\label{eq:Rxg}\\
R^{y} & = & \frac{F}{E_{o}}\cdot\frac{\lambda}{\pi d}\left[\frac{\alpha\beta}{\gamma}\, K^{x}-j\frac{ka}{2}\, K^{\theta}\right]\label{eq:Ryg}\\
R^{z} & = & -\frac{F}{E_{o}}\cdot\frac{\lambda}{\pi d}\left[\alpha K^{x}+j\frac{\beta}{\gamma}\cdot\frac{ka}{2}\, K^{\theta}\right]\label{eq:Rzg}\\
T^{x} & = & \alpha'+R^{x}\label{eq:Txg}\\
T^{y} & = & \beta'+\frac{F}{E_{o}}\cdot\frac{\lambda}{\pi d}\left[\frac{\alpha\beta}{\gamma}\, K^{x}+j\frac{ka}{2}\, K^{\theta}\right]\label{eq:Tyg}\\
T^{z} & = & \gamma'+\frac{F}{E_{o}}\cdot\frac{\lambda}{\pi d}\left[\alpha K^{x}-j\frac{\beta}{\gamma}\cdot\frac{ka}{2}\, K^{\theta}\right]\label{eq:Tzg}\end{eqnarray}

\noindent where we have set $z_{o}=0$ for simplicity.

Equations (\ref{eq:Rxg})-(\ref{eq:Tzg}) along with (\ref{eq:Kxg})-(\ref{eq:Ktg})
are the solution to the polarizing grid problem for cases where it
is assumed that $k'a\ll1$ and $a\ll d$. 

For predictions of measurements made in the laboratory, one merely
has to transform these coefficients to the laboratory coordinate system.
If we adopt for this system the coordinates of the incident/transmitted
$\left(u,v,w\right)$ and reflected $\left(u',v',w'\right)$ plane
waves defined in Figures \ref{fig:wire} and \ref{fig:u'v'w'}, the
last system of equations is simplified to:

\begin{eqnarray}
R^{u'} & = & -\frac{F}{E_{o}}\cdot\frac{\lambda}{\pi d}\cdot\frac{1}{\gamma\sqrt{1-\gamma^{2}}}\left[\beta K^{x}-j\alpha\gamma\,\frac{ka}{2}\, K^{\theta}\right]\label{eq:Ru}\\
R^{v'} & = & -\frac{F}{E_{o}}\cdot\frac{\lambda}{\pi d}\cdot\frac{1}{\gamma\sqrt{1-\gamma^{2}}}\left[\alpha\gamma K^{x}+j\beta\,\frac{ka}{2}\, K^{\theta}\right]\label{eq:Rv}\\
R^{w'} & = & 0\label{eq:Rw}\\
T^{u} & = & \alpha''-\frac{F}{E_{o}}\cdot\frac{\lambda}{\pi d}\cdot\frac{1}{\gamma\sqrt{1-\gamma^{2}}}\left[\beta K^{x}+j\alpha\gamma\,\frac{ka}{2}\, K^{\theta}\right]\label{eq:Tu}\\
T^{v} & = & \beta''-\frac{F}{E_{o}}\cdot\frac{\lambda}{\pi d}\cdot\frac{1}{\gamma\sqrt{1-\gamma^{2}}}\left[\alpha\gamma K^{x}-j\beta\,\frac{ka}{2}\, K^{\theta}\right]\label{eq:Tv}\\
T^{w} & = & 0\label{eq:Tw}\end{eqnarray}

\noindent with $\alpha''$ and $\beta''$ related to the incident
field by :

\[
\mathbf{E}_{_{i}}(\mathbf{r})=E_{o}\left(\alpha''\mathbf{e}_{u}+\beta''\mathbf{e}_{v}\right)\exp\left(-jkw\right)\,.\]

As can be seen, the reflected and transmitted fields have no component
along their respective direction of propagation as is required for
the propagation of plane waves in free space.

\subsection{Effects of grid imperfections.}

So far we have assumed that there were no imperfections in the construction
of the grid, obviously (and unfortunately) such is not the case in
a realistic situation. It would be instructive if we could calculate
the effects of errors that are likely to be introduced in the fabrication
process. In this section we will provide expressions that will allow
us to evaluate changes in the reflection and transmission coefficients
induced by two possible imperfections: random errors in wire spacing
and random variations in the size of the wire radius.

\subsubsection{Random errors in wire spacing.}

It is our experience that some of the commercially available grids
when observed under a microscope show some defects in their assembly.
Visually, the most obvious manifestation of this is inconsistency
in the spacing between wires. In order to calculate the effect of
these errors we have to go back to the discussion of section \ref{sec:grid}
that guided us into the evaluation of the induced current on the wires.
Since we can no longer assume that the wires are evenly spaced, we
must now realize that they will in general have different values for
the current and fields on their surface. This will be made more apparent
if we write down the expression for the $x$-component of the electric
field on the surface of the {}``center'' wire:

\begin{equation}
E_{g}^{x}(a)=-\left(1-\alpha^{2}\right)F\exp\left(-jk\left(\alpha x+\gamma z_{o}\right)\right)\sum_{n=-\infty}^{\infty}K_{n}^{x}\left(\xi\right)G_{n}\left(\xi\right)\label{eq:Exga}\end{equation}

\noindent with:

\begin{equation}
G_{n}\left(\xi\right)=\left\{ \begin{array}{ll}
H_{0}^{(2)}(k'a)\exp\left(-jk\beta\xi_{0}\right) & ,n=0\\
H_{0}^{(2)}\left(k'\left|nd+\xi_{n}\right|\right)\exp\left(-jk\beta\left(nd+\xi_{n}\right)\right) & ,n\neq0\,.\end{array}\right.\label{eq:Gn}\end{equation}

$K_{n}^{x}$ is the induced surface current on wire $n$ and the $\xi_{n}$
are statistically independent random errors in the positioning of
the wires. Now, if $E\left\{ x\right\} $ stands for the expected
value of $x$ and if we suppose that the errors have a zero mean,
we can write:\begin{eqnarray}
E\left\{ \xi_{m}^{r}\right\}  & = & 0\qquad\qquad\qquad\qquad\quad,r=1,3,5\ldots\label{eq:1c}\\
E\left\{ \xi_{m}^{r}\right\}  & = & E\left\{ \xi_{n}^{r}\right\} =E\left\{ \xi^{r}\right\} \qquad\,,\forall\, m,n\label{eq:2c}\\
E\left\{ \xi_{m}^{r}\xi_{n}^{s}\right\}  & = & E\left\{ \xi_{m}^{r}\right\} E\left\{ \xi_{n}^{s}\right\} \qquad\;\,\,\,\;,m\neq n\label{eq:3c}\\
\overline{\xi^{m}K^{x}} & = & \overline{\xi_{n}^{m}K_{n}^{x}}=E\left\{ \xi_{n}^{m}K_{n}^{x}\right\} \:\:\:,\forall\, m,n\,.\label{eq:4c}\end{eqnarray}

The first equation is deduced from the supposed evenness of the probability
density function of the errors, the second states that their statistics
are the same across the grid and the third expresses their statistical
independence. The last of these equations arises from the fact that
if we were to test a large number of similar grids, every wire would
exhibit the same average value $\overline{\xi^{m}K^{x}}$ for any
induced surface current moment (independent of its position $n$).

We will not go into the details of the calculations as they are somewhat
lengthy, but it can be shown that if we apply this last set of equations
and expand $K_{n}^{x}\left(\xi\right)$ and $G_{n}(\xi)$ with their
Taylor series around $\xi_{m}=0$ while solving for the boundary conditions,
we can find an expression (valid to the second order in $\xi$) for
the average longitudinal surface current:

\[
\overline{K^{x}}\simeq K^{x}\left[1-\frac{E\left\{ \xi^{2}\right\} }{\sum_{n=-\infty}^{\infty}G_{n}\left(0\right)}\sum_{m=-\infty}^{\infty}\left\{ \frac{1}{2}\frac{\partial^{2}G_{m}}{{\partial\xi_{m}}^{2}}-\frac{\left(1-\alpha^{2}\right)}{\Delta_{x}}\left[\frac{\partial G_{m}}{\partial\xi_{m}}\right]^{2}\right\} _{\xi_{m}=0}\right]\]

\noindent where $K^{x}$ is the current density induced on the wires
of a perfect grid and is given by equation (\ref{eq:Kxg}). One sees
that the errors bring a perturbation which is proportional to their
common variance.

If the same approach is used to calculate the effect of such random
errors on the value of the azimuthal surface current density $K_{n}^{\theta}$,
one finds that it remains unaffected:

\[
\overline{K^{\theta}}\simeq K^{\theta}\]

\noindent with $K^{\theta}$ given by equation (\ref{eq:Ktg}).

From this we could then proceed and calculate the expected value of
the reflection and transmission coefficients by evaluating equation
(\ref{eq:Exga}) (and the corresponding equations for the $y$ and
$z$ directions) in the far field, when this is accomplished we find
that the coefficients have exactly the same form as shown in equations
(\ref{eq:Rxg})-(\ref{eq:Tzg}) (or (\ref{eq:Ru})-(\ref{eq:Tw})).
We then merely have to replace $K^{x}$ and $K^{\theta}$ by $\overline{K^{x}}$
and $\overline{K^{\theta}}$ respectively.

\subsubsection{Random errors in the wire radius (wire to wire).}

Another type of error which can be analyzed is one concerning the
random variation in the size of the wire radius, which we will denote
by the letter $\eta$. More explicitly, we are considering differences
between wires and not variations along a single wire; we assume the
diameter of a wire to be constant but somewhat uncertain in its value.
This is the kind phenomenon that could occur if the wires were stretched
with slightly different tensions when installed or perhaps also in
cases where the wires have a finite ellipticity and are rotated between
rows. We can proceed in the same manner as we did in the last section
for the analysis of the boundary conditions and the fields away from
the grid. When this is done we get:

\begin{eqnarray*}
\overline{K^{x}} & \simeq & K^{x}\left[1-\frac{E\left\{ \eta^{2}\right\} }{\sum_{n=-\infty}^{\infty}G_{n}\left(0\right)}\left\{ \frac{1}{2}\frac{\partial^{2}G_{0}}{{\partial\eta_{0}}^{2}}-\frac{\left(1-\alpha^{2}\right)}{\Delta_{x}}\left[\frac{\partial G_{0}}{\partial\eta_{0}}\right]^{2}\right\} _{\eta_{0}=0}\right]\\
\overline{K^{\theta}} & \simeq & K^{\theta}\left[1-\frac{E\left\{ \eta^{2}\right\} }{Q_{0}\left(0\right)}\left\{ \frac{1}{2}\frac{\partial^{2}Q_{0}}{{\partial\eta_{0}}^{2}}+\frac{1}{a}\frac{\partial Q_{0}}{\partial\eta_{0}}-\frac{\sqrt{1-\alpha^{2}}}{\Delta_{\theta}}\left[\frac{\partial Q_{0}}{\partial\eta_{0}}\right]^{2}\right\} _{\eta_{0}=0}\right]\end{eqnarray*}

\noindent where $\eta_{n}$ is the random error in the size of the
radius of wire $n$, $G_{n}$ is given by equation (\ref{eq:Gn})
(with $\xi=0$ and $a$ replaced by $a+\eta$), $\Delta_{x}$ and
$\Delta_{\theta}$ by (\ref{eq:deltax}) and (\ref{eq:deltat}) respectively
and $Q_{0}\left(\eta_{0}\right)=H_{1}^{\left(2\right)}\left(k'\left(a_{0}+\eta_{0}\right)\right)$.
Again the expected value of the different coefficients can be obtained
by replacing the current components $K^{x}$ and $K^{\theta}$ by
$\overline{K^{x}}$ and $\overline{K^{\theta}}$ in equations (\ref{eq:Rxg})-(\ref{eq:Tzg})
(or (\ref{eq:Ru})-(\ref{eq:Tw})). It will also be noted that the
errors contain a perturbation term which is proportional to their
common variance.

\subsubsection{Predictions and comparison with experiments.}

Now that we have derived the equations for the reflection and transmission
coefficients it would be interesting to compare the predictions that
our model makes with experimental data. Although we have independently
treated the two types of errors, it is nevertheless obvious that within
the limit of precision of our analysis (small errors) that they can
both be simultaneously added in the expressions for the reflection
and transmission coefficients. Doing so would in principle allow us
to compare theory and experiments as actual grids are liable to exhibit
both kinds of defects. This also suggests though that it might be
impossible to separate the effects of both errors in measurements.
It turns out, however, that the perturbations caused by the errors
in the size of the wire are predicted by our model to be smaller than
those caused by the errors of the other type (for equivalent error
amplitudes), and we neglect them in the following comparison of theory
and measured grid properties.

\begin{figure}
\begin{center}\plotone{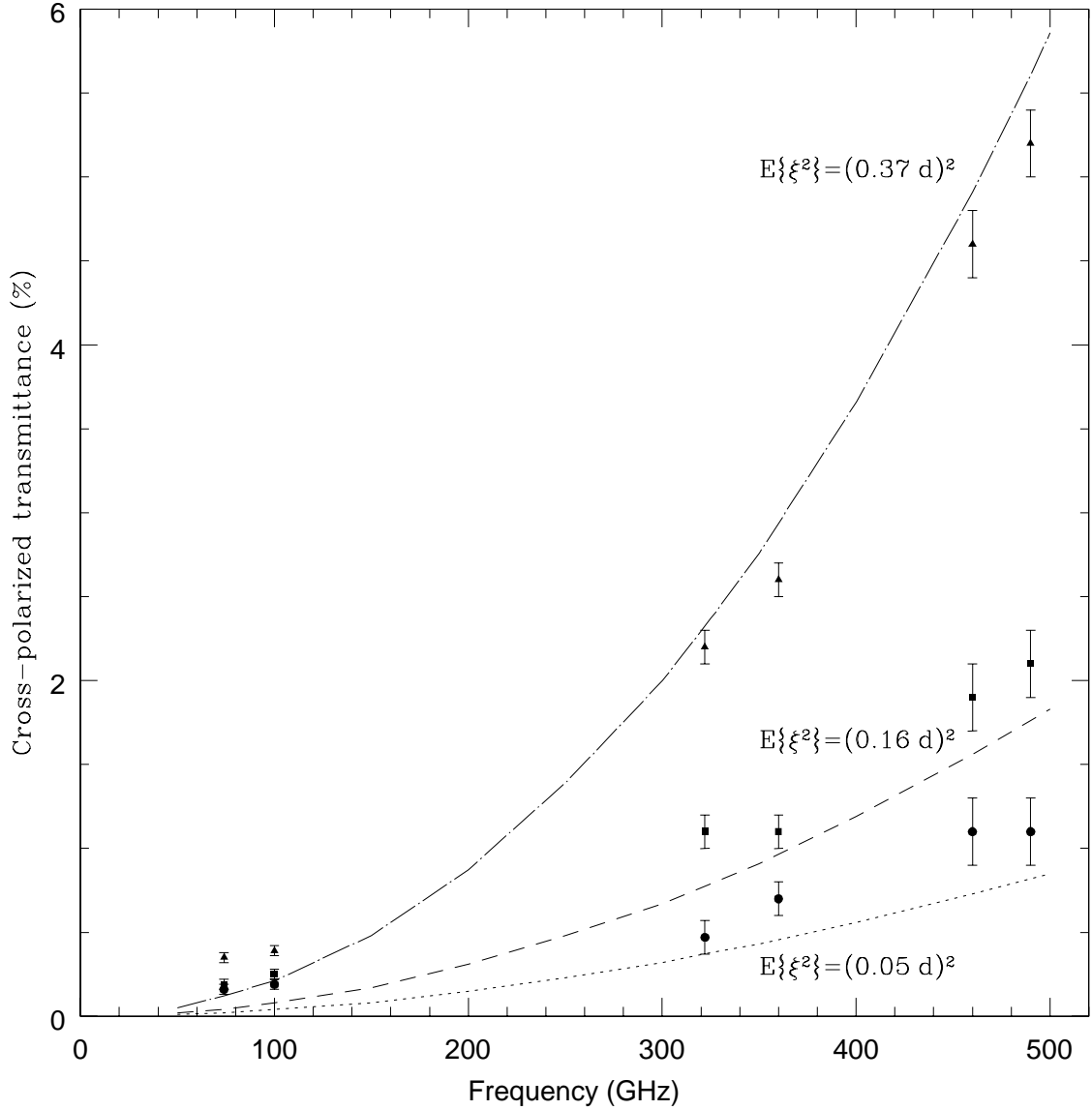}\end{center}

\caption{\label{fig:errors}Curves of predicted values for the cross-polarized
transmittance plotted against experimental data from \citet{Shapiro}
and \citet{Bloemhof 1998}. The three grids have a random error (1-$\sigma$)
in wire positioning of 5\%, 16\% and 37\% with mean distance between
wires of $103\,\mu$m, $109\,\mu$m and $114\,\mu$m respectively;
they all have a wire radius of $12.5\,\mu$m.}
\end{figure}

\citet{Shapiro} have published measurements of the unwanted cross-polarized
transmittance through three grids on which they had purposely introduced
random errors in the wire positioning. They quoted the errors in term
of the random variation in the distance between wires (pitch) with
amplitudes of 7\%, 23\% and 52\% of the mean wire separation (aimed
at $108\,\mu$m with a wire radius of $12.5\,\mu$m). We must divide
these values by a factor of $\sqrt{2}$ in order to relate them to
our errors $\xi_{n}$ since we have defined these as pertaining to
the absolute position of the wires. Figure \ref{fig:errors} shows
a comparison of our model's predictions with their measurements done
for cases where the incoming field is at normal incidence to the grid
and polarized parallel to the wire orientation. Although the agreement
is not perfect, the outcome is very satisfactory as the theoretical
curves exhibit the right behavior with frequency and error amplitude.

\subsection{The scattering matrix and the impedance model.\label{sec:imp}}

\subsubsection{The scattering matrix and the principal axes of a grid.\label{sec:axis}}

The relationship between the reflection and transmission coefficients
in equations (\ref{eq:Ru})-(\ref{eq:Tw}) is reminiscent of what
is often encountered in microwave engineering in the analysis of systems
that can be accurately dealt with using a lumped-elements model. With
this in mind, it is tempting to consider any problem involving a polarizing
grid by treating the different components as lumped and interconnected
through a transmission line of characteristic impedance $Z_{o}$ \citep{Lamb}.
We can then go ahead and model the grid as a $4$-port device since
the reflection and transmission coefficients given by the aforementioned
set of equations provides us with the scattering parameters at each
port.

In this context, it is more convenient to work with a single coordinate
system ($u,v,w$) (see Figures \ref{fig:wire} and \ref{fig:u'v'w'})
for both the incident/transmitted and reflected plane waves since
we can assume that their propagation is done along the same transmission
line (it is however understood that, in reality, away from normal
incidence the transmitted and reflected waves travel along different
axes). We therefore assume that the incident/transmitted fields travel
along the $w$-axis (with the $u$-axis vertical and the $v$-axis
horizontal) and the reflected fields along the negative $w$-axis
as seen from a given side of the grid. 

Since there are two possible independent states of polarization (with
the field aligned along the $u$ or $v$-axes), where the waves can
travel either toward or away from the grid, we need two ports on each
side of the grid. So for example, if the incident wave on a given
port has an electric field polarized along a given axis we can define
$4$ scattering parameters: one for the reflected signal at the input
port and three for the transmissions to the other ports. The same
thing can be done for every port leading to a total of  $16$ scattering
parameters.

In what follows, a scattering parameter $s_{mn}$ is defined with
the 3 ports $m\neq n$ terminated with the line characteristic impedance
$Z_{o}$. Also, each port $n$ has two signals: an incoming signal
$E_{n}^{+}$ and an outgoing signal $E_{n}^{-}$; $n=1,\,2$ ($3,\,4$
on the other side of the grid) refer to polarization along the $u$
and $v$ axes respectively. The scattering matrix relates the different
signals as follows:

\begin{equation}
\left[\begin{array}{c}
E_{1}^{-}\\
E_{2}^{-}\\
E_{3}^{-}\\
E_{4}^{-}\end{array}\right]=\left[\begin{array}{cccc}
s_{11} & s_{12} & s_{13} & s_{14}\\
s_{21} & s_{22} & s_{23} & s_{24}\\
s_{31} & s_{32} & s_{33} & s_{34}\\
s_{41} & s_{42} & s_{43} & s_{44}\end{array}\right]\left[\begin{array}{c}
E_{1}^{+}\\
E_{2}^{+}\\
E_{3}^{+}\\
E_{4}^{+}\end{array}\right]\,.\label{eq:smat}\end{equation}

The elements of the matrix can be directly evaluated from equations
(\ref{eq:Ru})-(\ref{eq:Tw}) and shown to be:

\begin{equation}
\mathbf{S}=\left[\begin{array}{cccc}
R^{uu} & R^{uv} & T^{uu} & T^{uv}\\
R^{uv} & R^{vv} & T^{uv} & T^{vv}\\
T^{uu} & T^{uv} & R^{uu} & R^{uv}\\
T^{uv} & T^{vv} & R^{uv} & R^{vv}\end{array}\right]\label{eq:Smatrix}\end{equation}

\noindent with:

\begin{eqnarray}
R^{uu} & = & -\frac{\lambda}{\pi d}\cdot\frac{1}{\gamma\left(1-\gamma^{2}\right)}\left[\beta^{2}\,\frac{N_{x}}{\Delta_{x}}-\alpha^{2}\gamma^{2}\,\frac{ka}{2}\cdot\frac{N_{\theta}}{\Delta_{\theta}}\right]\label{eq:Ruu}\\
R^{vv} & = & -\frac{\lambda}{\pi d}\cdot\frac{1}{\gamma\left(1-\gamma^{2}\right)}\left[\alpha^{2}\gamma^{2}\,\frac{N_{x}}{\Delta_{x}}-\beta^{2}\,\frac{ka}{2}\cdot\frac{N_{\theta}}{\Delta_{\theta}}\right]\label{eq:Rvv}\\
R^{uv} & = & -\frac{\lambda}{\pi d}\cdot\frac{\alpha\beta}{\left(1-\gamma^{2}\right)}\left[\frac{N_{x}}{\Delta_{x}}+\frac{ka}{2}\cdot\frac{N_{\theta}}{\Delta_{\theta}}\right]\label{eq:Ruv}\\
T^{uu} & = & 1-\frac{\lambda}{\pi d}\cdot\frac{1}{\gamma\left(1-\gamma^{2}\right)}\left[\beta^{2}\,\frac{N_{x}}{\Delta_{x}}+\alpha^{2}\gamma^{2}\,\frac{ka}{2}\cdot\frac{N_{\theta}}{\Delta_{\theta}}\right]\label{eq:Tuu}\\
T^{vv} & = & 1-\frac{\lambda}{\pi d}\cdot\frac{1}{\gamma\left(1-\gamma^{2}\right)}\left[\alpha^{2}\gamma^{2}\,\frac{N_{x}}{\Delta_{x}}+\beta^{2}\,\frac{ka}{2}\cdot\frac{N_{\theta}}{\Delta_{\theta}}\right]\label{eq:Tvv}\\
T^{uv} & = & -\frac{\lambda}{\pi d}\cdot\frac{\alpha\beta}{\left(1-\gamma^{2}\right)}\left[\frac{N_{x}}{\Delta_{x}}-\frac{ka}{2}\cdot\frac{N_{\theta}}{\Delta_{\theta}}\right]\label{eq:Tuv}\end{eqnarray}

\noindent where $N_{x}$, $\Delta_{x}$, $N_{\theta}$ and $\Delta_{\theta}$
are given by equations (\ref{eq:Nx}), (\ref{eq:deltax}), (\ref{eq:Nt})
and (\ref{eq:deltat}) respectively. 

We can go one step further and render things considerably simpler
if we make a change of coordinates and use the following as eigenvectors
instead of $\mathbf{e}_{\mathbf{u}}$ and $\mathbf{e}_{\mathbf{v}}$:

\begin{eqnarray*}
\mathbf{p}_{1} & = & \frac{\beta\mathbf{e}_{\mathbf{u}}+\alpha\gamma\mathbf{e}_{\mathbf{v}}}{\sqrt{\beta^{2}+\alpha^{2}\gamma^{2}}}\\
\mathbf{p}_{2} & = & \frac{-\alpha\gamma\mathbf{e}_{\mathbf{u}}+\beta\mathbf{e}_{\mathbf{v}}}{\sqrt{\beta^{2}+\alpha^{2}\gamma^{2}}}\,.\end{eqnarray*}

From now on we will refer to these as the \textit{principal axes}
of the grid (for reasons that will soon become apparent). A close
examination of the first of these two equations shows that $\mathbf{p}_{1}$
is parallel to the projection of the direction of the wires in the
plane of the incident field. The matrix $\mathbf{S}$ then takes a
simpler \emph{}form (we also interchange the second row with the third
and the second column with the third): 

\begin{equation}
\mathbf{S}=\left[\begin{array}{cccc}
R_{\Vert} & T_{\Vert} & 0 & 0\\
T_{\Vert} & R_{\Vert} & 0 & 0\\
0 & 0 & R_{\bot} & T_{\bot}\\
0 & 0 & T_{\bot} & R_{\bot}\end{array}\right]\label{eq:Sd}\end{equation}

\noindent where

\begin{eqnarray}
R_{\Vert} & = & -\frac{\lambda}{\pi d}\cdot\frac{\left(1-\alpha^{2}\right)}{\gamma}\cdot\frac{N_{x}}{\Delta_{x}}\label{eq:Rp}\\
R_{\bot} & = & \frac{\left(1-\alpha^{2}\right)}{\gamma}\cdot\frac{a}{d}\cdot\frac{N_{\theta}}{\Delta_{\theta}}\label{eq:Rn}\\
T_{\Vert} & = & 1+R_{\Vert}\label{eq:Tp}\\
T_{\bot} & = & 1-R_{\bot}\,.\label{eq:Tn}\end{eqnarray}

We then have a further simplification in the modeling of the grid,
evidently equations (\ref{eq:Rp})-(\ref{eq:Tn}) represent the reflection
and transmission coefficients along the two principal axes.

This last representation has the advantage of simplifying calculations
since it allows us to decompose any incident field into two non-interacting
components, one along each one of the principal axes. That is, a field
polarized along one of the principal axes scatters only in this same
polarization state (as can be deduced from the block-diagonal form
of equation (\ref{eq:Sd})). It is also interesting to note that even
though we have defined the principal axes within the framework of
our approximation of the grid ($k'a\ll1$ and $a\ll d$), the result
obtained here still holds in the general case (see Appendix A for
a proof). This implies that for the case where one wishes to use a
different approach to solve (numerically or otherwise) the scattering
off a grid of arbitrary characteristics, it will always be possible
to split the incoming field along the principal axes therefore avoiding
cross-polarization terms and greatly simplifying the solution.

\subsubsection{The impedance model.}

It seems reasonable to think that a grid could also be modeled with
another representation where the scattering matrix is replaced by
an impedance matrix which contains the same number of elements since
as before, the grid is still treated as a 4-port device. In this case
however, the matrix relates the total voltages (electric fields) and
currents (magnetic fields) between each and every port \citep{Collin}.
The scattering matrix formulation follows more naturally from our
analysis and has the advantage of dealing with quantities (reflection
and transmission coefficients) which are directly measurable whereas
impedances are not (at least at the wavelengths considered here).
The impedance model has however received a great deal of attention
in the literature and often seems to be the way in which polarizing
grids are characterized \citep{Wait 1954,Wait 1955a, Larsen}. 

Taking advantage of the principal axes representation, it is possible
to treat each two-dimensional block of the scattering matrix (equation
(\ref{eq:Sd})) separately. It can be shown that the impedance matrix
$\mathbf{Z}_{b}$ corresponding to a given block $\mathbf{S}_{b}$
can be expressed as: 

\begin{equation}
\mathbf{Z}_{b}=Z_{o}\cdot\left(\mathbf{I}\pm\mathbf{S}_{b}\right)\cdot\left(\mathbf{I}\mp\mathbf{S}_{b}\right)^{-1}\label{eq:Z}\end{equation}

\noindent where $\mathbf{I}$ is the unit matrix and the upper and
lower signs correspond respectively to the upper left and lower right
blocks of the scattering matrix (equation (\ref{eq:Sd})). Applying
this last equation to equation (\ref{eq:Sd}) we get:

\[
\mathbf{Z}=\left[\begin{array}{cccc}
Z_{p} & Z_{p} & 0 & 0\\
Z_{p} & Z_{p} & 0 & 0\\
0 & 0 & Z_{n} & -Z_{n}\\
0 & 0 & -Z_{n} & Z_{n}\end{array}\right]\]

\noindent with: \begin{eqnarray*}
Z_{p} & = & -\frac{Z_{o}}{2}\cdot\frac{1+R_{\Vert}}{R_{\Vert}}\\
Z_{n} & = & \frac{Z_{o}}{2}\cdot\frac{1-R_{\bot}}{R_{\bot}}\,.\end{eqnarray*}

It follows quite naturally from equations (\ref{eq:Rp})-(\ref{eq:Tn})
that we could have defined two impedances $Z_{\Vert}$ and $Z_{\bot}$:

\begin{eqnarray*}
Z_{\Vert} & = & Z_{o}\cdot\frac{1+R_{\Vert}}{1-R_{\Vert}}\\
Z_{\bot} & = & Z_{o}\cdot\frac{1-R_{\bot}}{1+R_{\bot}}\,.\end{eqnarray*}

One can easily verify that $Z_{\Vert}$ and $Z_{\bot}$ are respectively
equal to $Z_{p}$ and $Z_{n}$ placed in parallel to the cha\-rac\-te\-ristic
im\-pe\-dance $Z_{o}$. We therefore see that the impedance matrix
gives the {}``actual'' impedance of the grid along each of the principal
axes whereas the scattering matrix includes, as should be expected,
the contribution of the loads of characteristic impedance $Z_{o}$
which is assumed to be connected to the appropriate ports when defining
its parameters.

\subsection{Approximations and selection of a grid.\label{sec:approx}}

We will now study more closely our simpler equations (\ref{eq:Rp})
and (\ref{eq:Rn}) for the reflection coefficients $R_{\Vert}$ and
$R_{\bot}$and try to find relations that will allow us to find a
set of optimum parameters for the selection of a grid. But before
we do so, it will be to our advantage to approximate the expressions
for $\Delta_{x}$ and $\Delta_{\theta}$ (equations (\ref{eq:deltax})
and (\ref{eq:deltat})). 

So if we limit ourselves to situations where $d\ll\lambda$, $a\ll\lambda$
and $Z_{s}\ll Z_{o}$ (good conducting wires) and use the proper expansion
for Hankel's functions and series of Hankel's functions applicable
in such cases (small arguments) we find:

\begin{eqnarray}
\Delta_{x} & \simeq & \left(1-\alpha^{2}\right)\left[\left\{ \frac{\lambda}{\pi\gamma d}-\left(\frac{k'a}{2}\right)^{2}+\frac{1}{\sqrt{\left(1-\alpha^{2}\right)\pi Z_{o}\sigma\lambda}}\left(\frac{2}{k'a}\right)\right\} \right.\nonumber \\
 &  & +\left.j\frac{2}{\pi}\left\{ \ln\left(\frac{d}{2\pi a}\right)+\frac{\pi^{2}}{6}\left(\frac{d\gamma}{\lambda}\right)^{2}-\left(\frac{k'a}{2}\right)^{2}\left[1-\Psi-\ln\left(\frac{k'a}{2}\right)\right]\right.\right.\nonumber \\
 &  & +\left.\left.\rule{0in}{4ex}\sqrt{\frac{\pi}{4\left(1-\alpha^{2}\right)Z_{o}\sigma\lambda}}\left(\frac{2}{k'a}\right)\right\} \right]\label{eq:adeltx}\\
\Delta_{\theta} & \simeq & -\left(1-\alpha^{2}\right)\,\frac{Z_{s}}{Z_{o}}\cdot\frac{2}{\pi}\,\ln\left(\frac{d}{2\pi a}\right)+j\left\{ \frac{\lambda}{\pi^{2}a}+\left(1-\alpha^{2}\right)\,\frac{Z_{s}}{Z_{o}}\cdot\frac{\lambda}{\pi\gamma d}\right\} \label{eq:adelt}\end{eqnarray}

\noindent where $\Psi\simeq0.577215$ is Euler's constant. In equation
(\ref{eq:adelt}) we have kept things to the lowest order possible
and we did not expand $Z_{s}$; the same is not true for equation
(\ref{eq:adeltx}) for reasons that we shall encounter shortly.

We turn now to the problem of selecting the right parameters for a
grid. If we decompose a given incident field into two components along
the principal axes $\mathbf{p}_{1}$ and $\mathbf{p}_{2}$ (see section
\ref{sec:axis}), a perfect grid would completely reflect the first
of these and transmit the second ($R_{\Vert}=-1$ and $T_{\bot}=1$).
As we will soon see, the coefficient of reflection $R_{\bot}$ is
proportional to $\frac{a^{2}}{d\lambda}$ when $Z_{s}\rightarrow0$
and is therefore a very small quantity for the cases considered here
and we will not worry about it anymore (i.e., $T_{\bot}$ is nearly
equal to unity). The condition of total reflection will dictate our
choice for the parameters of the grid. A close study of equation (\ref{eq:Rp})
tells us that in order to achieve perfect reflection we must simultaneously
satisfy the following relations for the real and imaginary parts of
$\Delta_{x}$ (for what follows we assume $N_{x}\simeq1$, see equation
(\ref{eq:Nx})):

\begin{eqnarray*}
\mbox{Re}\left\{ \Delta_{x}\right\}  & = & \frac{\lambda}{\pi d}\cdot\frac{\left(1-\alpha^{2}\right)}{\gamma}\\
\mbox{Im}\left\{ \Delta_{x}\right\}  & = & 0\,.\end{eqnarray*}

Solving for these we then get:

\begin{eqnarray}
a & \simeq & \left[\frac{\lambda^{5}}{\left(1-\alpha^{2}\right)^{4}\pi^{7}\sigma Z_{o}}\right]^{\frac{1}{6}}\label{eq:opta}\\
d & \simeq & 2\pi a\,.\label{eq:optd}\end{eqnarray}

Had we kept equation (\ref{eq:adeltx}) to the lowest order, we would
have been unable to specify an optimum value for the wire radius but
only the relation that binds $d$ to $a$. It is also of interest
to note that for a given wavelength, the finite size of the wire radius
is, to this level of approximation, dictated by the conductivity $\sigma$;
if we let $\sigma\rightarrow\infty$ then there is no restriction
on the smallness of the radius. 

In a quantitative example to demonstrate the values that can be expected
for $a$ and $d$, assume that we are working at normal incidence
at a wavelength of 1 mm with a grid made of copper ($\sigma=5.8\times10^{7}$
$\Omega^{-1}$m$^{-1}$). Using these, we obtain $a\simeq11\,\mu$m
and $d\simeq70\,\mu$m.

At this point it is appropriate to discuss the implications of the
two assumptions we made at the beginning concerning the wire radius
and spacing, namely that $k'a\ll1$ and $a\ll d$. It is important
to make sure that a given choice of grid parameters are well within
the boundaries of applicability of our model. As a means of determining
these boundaries, we simulated the response of grids (and assemblies
of grids, see section \ref{sec:RP}) for different combinations of
wire radius and spacing and made sure that the results obtained were
reliable (for example, it is obviously imperative that the magnitude
of the reflection and transmission coefficients never exceed unity).
As it turns out, there is a fairly strong restriction linking the
size of the wires and the wavelength, but if one makes sure that $\lambda>40\, a$
then one seems to be well within safe modeling conditions. $a$ cannot
be too small either. However, since for a good conductor (again let's
use copper) the skin depth at 1 mm is on the order of $0.1\,\mu$m,
our assumption of the existence of an idealized surface current is
more than adequate. It seems that the second restriction concerning
the spacing of the wires is not as binding as the first one. It is
clear that $d>2\, a$ for if not the wires would be touching, but
it appears that everything is fine for $d>4\, a$. Our proposed optimized
values for the grid are therefore justified. 

It is also appropriate to point out that using equations (\ref{eq:adeltx})
and (\ref{eq:adelt}) for $\Delta_{x}$ and $\Delta_{\theta}$ (with
or without the optimal values for $a$ and $d$ given by equations
(\ref{eq:opta}) and (\ref{eq:optd})) along with equations (\ref{eq:Rp})-(\ref{eq:Tn})
for the reflection and transmission coefficients along the principal
axes renders the task of calculating the response of a grid a rather
simple one. It becomes unnecessary to confront the more intimidating
representations derived earlier in section \ref{sec:grid} (compare
with equations (\ref{eq:Kxg})-(\ref{eq:Tzg})). For example, to the
lowest order, we get for the reflection coefficients:

\begin{eqnarray}
R_{\Vert} & \simeq & \frac{-1}{1+j\frac{2\gamma d}{\lambda}\ln\left(\frac{d}{2\pi a}\right)}\label{eq:appRpara}\\
R_{\bot} & \simeq & -j\frac{\left(1-\alpha^{2}\right)}{\gamma}\cdot\frac{\pi^{2}a^{2}}{\lambda d}\label{eq:appRperp}\end{eqnarray}

\noindent which are in agreement with known results \citep{Larsen}
(more precisely, for the case of normal incidence discussed in \citet{Larsen},
equation (\ref{eq:appRpara}) reduces to the result presented there
whereas equation (\ref{eq:appRperp}) differs by a factor of two or
three depending on which approximation it is compared to).

\section{The reflecting polarizer. }

\label{sec:RP}

\subsection{Analysis.\label{sec:analysis}}

From the solution of the polarizing grid it is a somewhat natural
extension to consider the more complicated problem of the reflecting
polarizer. A reflecting polarizer consists of an assembly where a
polarizing grid, like the one studied in the last section, is followed
by a mirror paralleling it at some distance $z_{o}$ behind (effectively
placing the mirror at $z=0$).

It is appropriate in this case to use the method of images to solve
this problem \citep{Wait 1954}. We then assume that images of both
the incident and scattered fields are emanating from the other side
of the mirror. This is equivalent to saying that the image world is
made of a grid positioned at $z=-z_{o}$ with an image incident field
impinging on it. Assuming that the mirror is made of a material of
good conductivity, one can write for the image incident field $\mathbf{E}_{i}'(\mathbf{r})$:

\[
\mathbf{E}_{i}'(\mathbf{r})=E_{o}\left(\alpha_{m}'\mathbf{e}_{x}+\beta_{m}'\mathbf{e}_{y}-\gamma_{m}'\mathbf{e}_{z}\right)\exp(-jk(\alpha x+\beta y-\gamma z))\]

\noindent with:

\begin{eqnarray}
\alpha_{m}' & = & \frac{1}{\left(1-\gamma^{2}\right)}\left[\alpha'\left(\alpha^{2}R_{TM}+\beta^{2}R_{TE}\right)+\alpha\beta\beta'\left(R_{TM}-R_{TE}\right)\right]\label{eq:alppm}\\
\beta_{m}' & = & \frac{1}{\left(1-\gamma^{2}\right)}\left[\beta'\left(\alpha^{2}R_{TE}+\beta^{2}R_{TM}\right)+\alpha\beta\alpha'\left(R_{TM}-R_{TE}\right)\right]\label{eq:betpm}\\
\gamma_{m}' & = & \gamma'R_{TM}\label{eq:gampm}\end{eqnarray}

\noindent where $R_{TE}$ and $R_{TM}$ are the reflection coefficients
of the mirror, with a dependency on the angle of incidence, for transverse
electric and transverse magnetic modes of incoming radiation respectively
\citep{Fowles}. It is important to note that these tranverse modes
of radiation are not the same as those introduced in section \ref{sec:wire},
they are defined here in relation to the plane which is parallel the
normal vector out of the surface of the mirror ($-\mathbf{e}_{z}$)
and the wave vector $\mathbf{k}$.

One can go through calculations similar to those carried out in section
\ref{sec:grid} and find the following relations between the components
of the total surface current densities of the {}``real'' and image
grids:

\begin{eqnarray*}
K'^{x} & = & \frac{\alpha_{m}'}{\alpha'}\, K^{x}\\
K'^{\theta} & = & -\frac{\gamma_{m}'\beta-\beta_{m}'\gamma}{\gamma'\beta-\beta'\gamma}\, K^{\theta}\end{eqnarray*}

\noindent where $\mathbf{K}'$ stands for the surface current of the
image grid.

From these and by matching the boundary conditions at the grid, it
is straightforward, but tedious, to solve for the problem. We give
here the final results:

\begin{eqnarray}
K^{x} & = & \frac{E_{o}}{F}\cdot\frac{2j\alpha'\,\mbox{sn}(k\gamma h)\, N_{x}}{\left(1-\alpha^{2}\right)\Delta S_{1}-j\frac{Z_{s}}{Z_{o}}\sqrt{1-\alpha^{2}}\Delta S_{2}}\exp\left(-jk\gamma h\right)\label{eq:Kxp}\\
K^{\theta} & = & \frac{E_{o}}{F}\cdot\frac{-2j\left(\gamma'\beta-\beta'\gamma\right)\,\mbox{cs}(k\gamma h)\, N_{\theta}}{\sqrt{1-\alpha^{2}}\Sigma S_{2}+j\frac{Z_{s}}{Z_{o}}\left(1-\alpha^{2}\right)\Sigma S_{1}}\exp\left(-jk\gamma h\right)\label{eq:Ktp}\\
R^{x} & = & \alpha_{m}'-\frac{F}{E_{o}}\cdot\frac{2\lambda}{\pi d}\cdot\frac{\left(1-\alpha^{2}\right)}{\gamma}\, j\,\mbox{sn}(k\gamma h)\, K^{x}\exp\left(jk\gamma h\right)\label{eq:Rxp}\\
R^{y} & = & \beta_{m}'+\frac{F}{E_{o}}\cdot j\frac{2\lambda}{\pi d}\left[\frac{\alpha\beta}{\gamma}\,\mbox{sn}(k\gamma h)\, K^{x}-\frac{ka}{2}\,\mbox{cs}(k\gamma h)\, K^{\theta}\right]\exp\left(jk\gamma h\right)\label{eq:Ryp}\\
R^{z} & = & -\gamma_{m}'-\frac{F}{E_{o}}\cdot j\frac{2\lambda}{\pi d}\left[\alpha\,\mbox{sn}(k\gamma h)\, K^{x}+\frac{\beta}{\gamma}\cdot\frac{ka}{2}\,\mbox{cs}(k\gamma h)\, K^{\theta}\right]\exp\left(jk\gamma h\right)\label{eq:Rzp}\end{eqnarray}

\noindent where

\begin{eqnarray*}
\mbox{sn}(x) & = & \frac{1}{2j}\left[\exp(jx)-r_{x}\exp(-jx)\right]\\
\mbox{cs}(x) & = & \frac{1}{2}\left[\exp(jx)+r_{\theta}\exp(-jx)\right]\\
\Delta S_{1} & = & H_{0}^{(2)}(k'a)-r_{x}H_{0}^{(2)}(k'2h)\\
 &  & +2\sum_{n=1}^{\infty}\left[H_{0}^{(2)}(k'nd)-r_{x}H_{0}^{(2)}\left(k'\sqrt{(nd)^{2}+4h^{2}}\right)\right]\cos(k\beta nd)\\
\Delta S_{2} & = & H_{1}^{(2)}(k'a)-r_{x}H_{1}^{(2)}(k'2h)\\
\Sigma S_{1} & = & H_{0}^{(2)}(k'a)+r_{\theta}H_{0}^{(2)}(k'2h)\\
 &  & +2\sum_{n=1}^{\infty}\left[H_{0}^{(2)}(k'nd)+r_{\theta}H_{0}^{(2)}\left(k'\sqrt{(nd)^{2}+4h^{2}}\right)\right]\cos(k\beta nd)\\
\Sigma S_{2} & = & H_{1}^{(2)}(k'a)+r_{\theta}H_{1}^{(2)}(k'2h)\end{eqnarray*}

\noindent and \begin{eqnarray*}
r_{x} & = & -\frac{\alpha_{m}'}{\alpha'}\\
r_{\theta} & = & -\frac{\gamma_{m}'\beta-\beta_{m}'\gamma}{\gamma'\beta-\beta'\gamma}\,.\end{eqnarray*}

We have also replaced $-z_{o}$ by $h$ ($h>0$) so that the distance
between the mirror and the grid is expressed by a positive quantity.
This set of equations along with equations (\ref{eq:alppm})-(\ref{eq:gampm})
give us the solution of the reflecting polarizer problem for cases
where $k'a\ll1$ and $a\ll d$. 

As was the case for the polarizing grid, if we transform those coefficients
to the laboratory frame of coordinates ($u'$, $v'$, $w'$) (see
Figure \ref{fig:u'v'w'}) we obtain:

\begin{eqnarray}
R^{u'} & = & \alpha_{m}''-\frac{F}{E_{o}}\cdot\frac{2\lambda}{\pi d}\cdot\frac{j\exp\left(jk\gamma h\right)}{\gamma\sqrt{1-\gamma^{2}}}\cdot\left[\beta\,\mbox{sn}(k\gamma h)\, K^{x}-\alpha\gamma\,\frac{ka}{2}\,\mbox{cs}\left(k\gamma h\right)\, K^{\theta}\right]\label{eq:Rup}\\
R^{v'} & = & \beta_{m}''-\frac{F}{E_{o}}\cdot\frac{2\lambda}{\pi d}\cdot\frac{j\exp\left(jk\gamma h\right)}{\gamma\sqrt{1-\gamma^{2}}}\cdot\left[\alpha\gamma\,\mbox{sn}(k\gamma h)\, K^{x}+\beta\,\frac{ka}{2}\,\mbox{cs}(k\gamma h)\, K^{\theta}\right]\label{eq:Rvp}\\
R^{w'} & = & 0\label{eq:Rwp}\end{eqnarray}

\noindent with $\alpha_{m}''$ and $\beta_{m}''$ given by :

\begin{eqnarray}
\alpha_{m}'' & = & \frac{1}{\left(1-\gamma^{2}\right)}\left[\alpha''\left(\alpha^{2}R_{TM}+\beta^{2}R_{TE}\right)+\alpha\beta\beta''\left(R_{TM}-R_{TE}\right)\right]\label{eq:alpppm}\\
\beta_{m}'' & = & \frac{1}{\left(1-\gamma^{2}\right)}\left[\beta''\left(\alpha^{2}R_{TE}+\beta^{2}R_{TM}\right)+\alpha\beta\alpha''\left(R_{TM}-R_{TE}\right)\right]\label{eq:betppm}\end{eqnarray}

\noindent and finally $\alpha''$ and $\beta''$ are related to the
incident field by:

\[
\mathbf{E}_{_{i}}(\mathbf{r})=E_{o}\left(\alpha''\mathbf{e}_{u}+\beta''\mathbf{e}_{v}\right)\exp\left(-jkw\right)\,.\]

\subsection{Solution using the scattering matrix.}

The scattering matrix representation of the polarizing grid gives
us the advantage of rendering possible the solution of problems that
would be otherwise extremely difficult, if not impossible, to solve
using Maxwell's equations. For example, a solution of the reflecting
polarizer problem is straightforward if we {}``connect'' the mirror
to ports $3$ and $4$ of the grid at a distance $h$ behind. Using
the definitions introduced in the discussion leading to equation (\ref{eq:smat})
we have:

\begin{eqnarray*}
E_{3}^{+} & = & E_{3}^{-}R_{TE}\exp(-jk\gamma2h)\\
E_{4}^{+} & = & E_{4}^{-}R_{TM}\exp(-jk\gamma2h)\end{eqnarray*}

\noindent where $R_{TE}$ and $R_{TM}$ are as defined in the previous
section.

We can then solve for $E_{1}^{-}$ and $E_{2}^{-}$ and find results
that are in agreement with those obtained in the previous section.

\subsection{Experimental results.\label{sec:exp}}

Reflecting polarizers like those studied here were tested at the OVRO
for polarimetry in the wavelength ranges of 1.3 mm as well as successfully
used for polarimetry observations at 3 mm \citep{Akeson 1997, Akeson 1996}.
They are composed of an aluminum mirror and a grid of gold-plated
tungsten wires of $25\,\mu$m diameter and spaced at an interval of
$125\,\mu$m. The inside diameter of the grid is roughly 16 cm, some
25 times bigger than the incident Gaussian beam at 1.3 mm \citep{Akeson 1997}.

In this section, we will compare data obtained in the calibration
of these polarizers at 1.3 mm with the model calculated earlier. In
the experimental set-up, the incident beam is composed of radiation
emanating from a hot load (absorber at room temperature) polarized
along the vertical axis and a cold load (absorber in liquid nitrogen)
along the horizontal axis. The beam is incident on the polarizers
at an angle of $\chi_{i}=34\deg$. with the grid rotated by $\varphi_{g}=\pm50.3\deg$.
relative to the vertical, all in the coordinate system of the laboratory
(coordinates ($u$,$v$,$w$) of Figure \ref{fig:wire}). These values
can be inserted in the appropriate equations of our earlier analysis
and used to test our model against the experimental data. 

The calibration consists in using our model to map out the actual
distance between the grid and the mirror as the latter is moved with
a micro-positioner which is part of the assembly. When this is done,
this distance can then be precisely adjusted to $\frac{\lambda}{8\gamma}$
in order to use the polarizer as a reflecting quarter-wave plate for
polarimetry measurements. The grid rotation angle $\varphi_{g}$ must
also be calibrated so that it can be set to the proper value that
will allow the transformation of incident linear polarization to circular
polarization. (This condition is met for $\beta=\pm\alpha\gamma$
(or $\tan(\varphi_{g})=\frac{\pm1}{\cos(\chi_{i})}$), as can be asserted
from our earlier discussion of the principal axes of a grid in section
\ref{sec:imp}; this gives $\varphi_{g}=\pm50.3\deg$ as quoted above).

\begin{figure}
\begin{center}\plotone{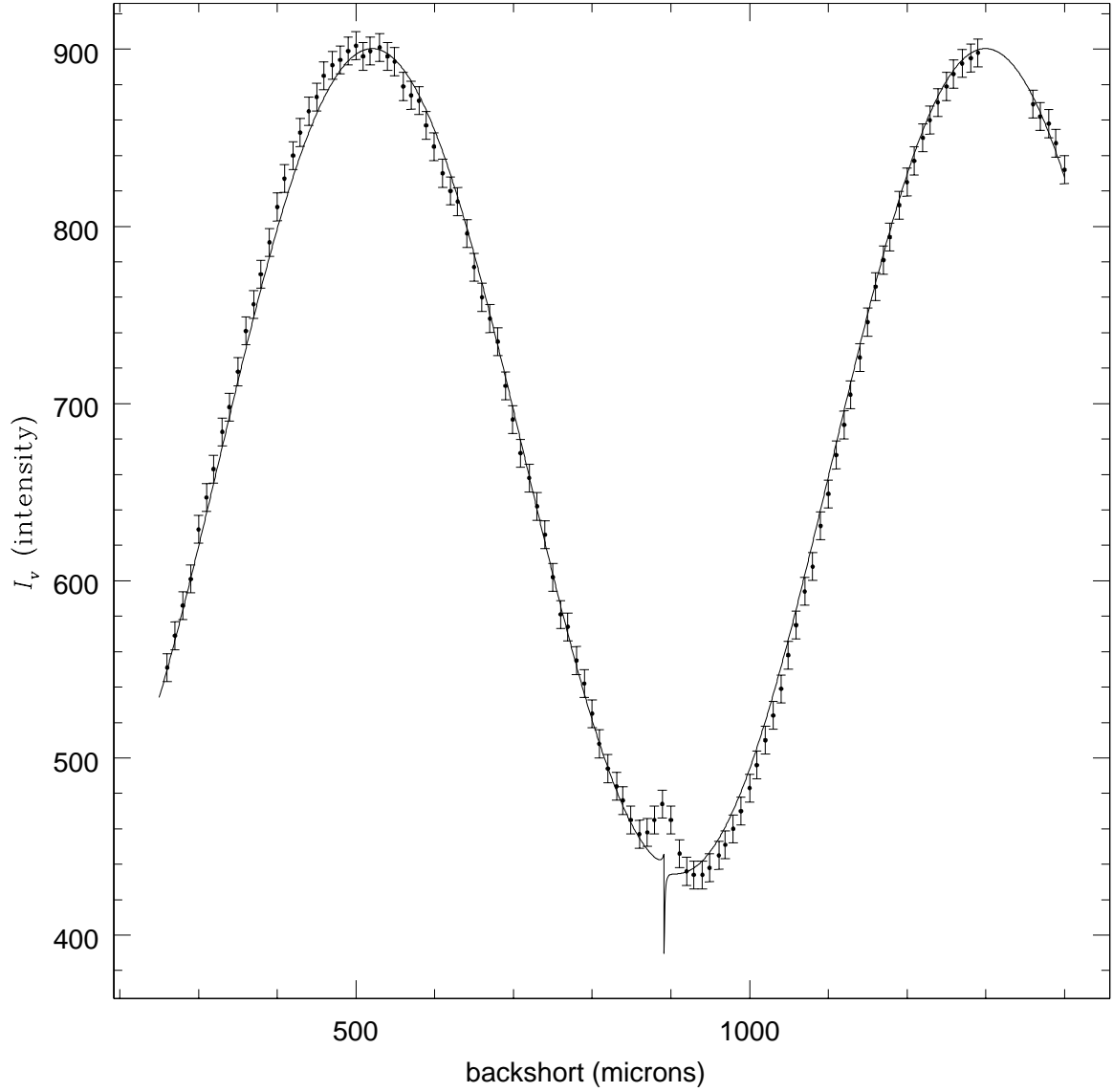}\end{center}

\caption{\label{fig:results}Calibration data from antenna \#6 of the OVRO
array. Data points are shown with an error bar and the solid curve
is a least squares fit from the model presented in this paper. The
intensity is in arbitrary units.}
\end{figure}

Figure \ref{fig:results} shows the results obtained from such measurements
(of the reflected polarized intensity along the horizontal $v'$-axis
in the laboratory coordinate system) made on antenna \#6 of the OVRO
array at a frequency of 232.037 GHz when the separation between the
grid and the mirror is varied through a range of a several hundreds
of microns. Accompanying the data points is a least square fit of
our model (solid curve) with no free parameters as far as the grid
is concerned, only the hot and cold load levels and the offset in
the mirror-grid separation were allowed to be fitted. The agreement
is very good. The main shortcoming of the fit is at a backshort position
of roughly $900\,\mu$m where a resonance is evident from the data.
The model also shows a resonance at the same position but the fit
is not perfect. This feature is caused by the small amount of unwanted
transmission from the component of the incident electric field aligned
with one of the principal axes ($\mathbf{p}_{1}$) which gets trapped
between the grid and the mirror.

Before we try to explain the differences in width and shape of the
resonance, we would first like to show two ways by which it can be
suppressed (such a response from the polarizer is a nuisance when
trying to calibrate it and should be avoided).

\begin{figure}
\begin{center}\plotone{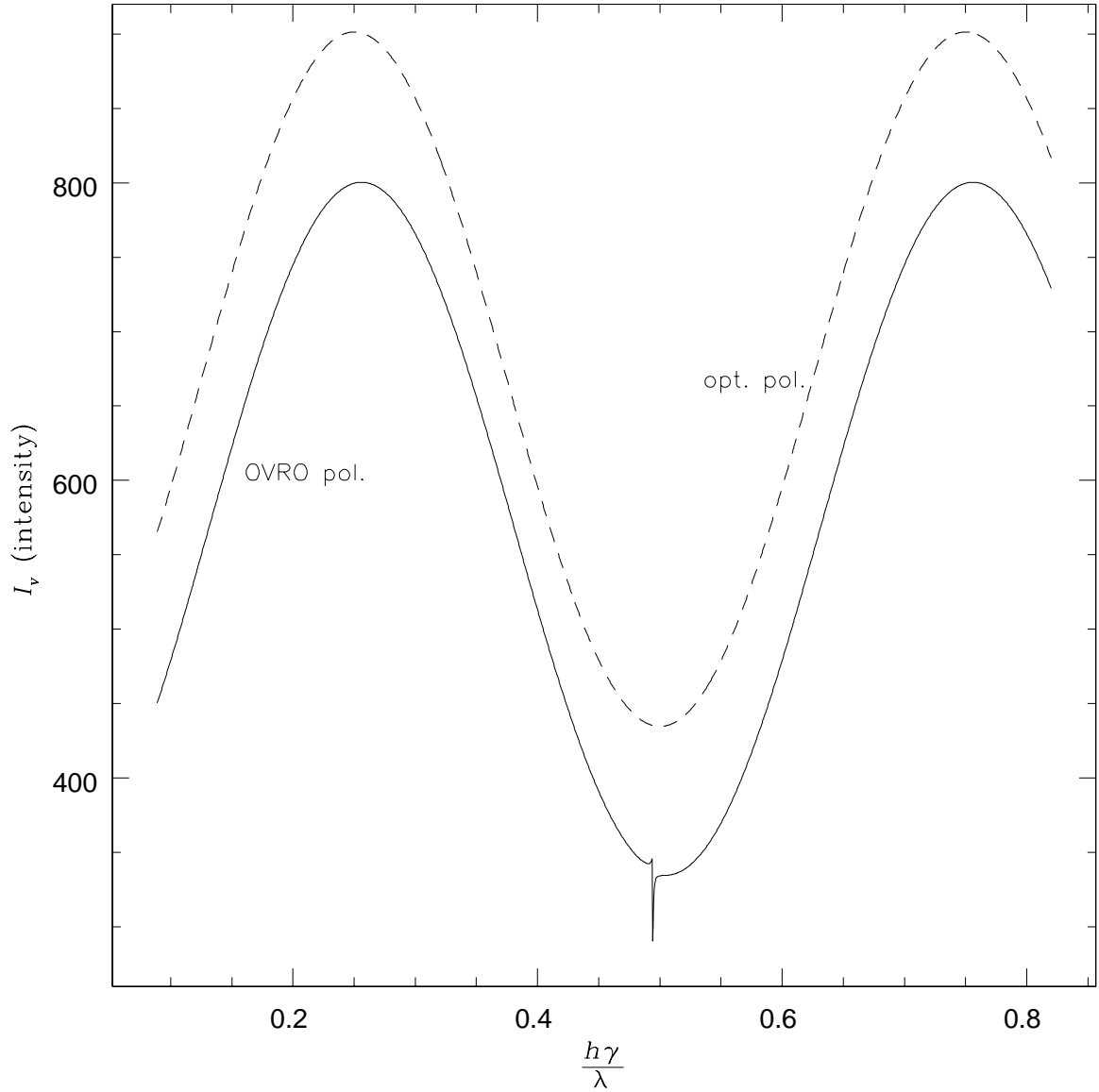}\end{center}

\caption{\label{fig:comp}Comparison of the predicted results obtained for
the polarizer tested at OVRO (solid curve) and our optimized polarizer
(broken curve). The resonance is not present on the optimized polarizer's
response. The two curves are plotted with a small vertical offset
between them. The intensity is in arbitrary units.}
\end{figure}

First, reducing the amount of unwanted transmission through the grid
would certainly have a damping effect on the resonance. We have shown
how to do just that in section \ref{sec:approx} when defining a set
of optimum parameters for a grid, so using equations (\ref{eq:opta})
and (\ref{eq:optd}) we find for our application $a\simeq24\,\mu$m
and $d\simeq148\,\mu$m. Figure \ref{fig:comp} shows a comparison
of the simulated responses for the polarizer tested at OVRO and our
optimized polarizer. As can be seen, any sign of the resonance has
disappeared in the latter. 

Another way of avoiding the resonance, while still using the same
original grid with $a=12.5\,\mu$m and $d=125\,\mu$m, is to replace
the mirror with another grid \citep{Young} and rotate both of them
in such a way that the projected orientation of their wires in the
plane of the incident field are aligned with a principal axis. By
this we mean that the two grids have their angle of rotation specified
by $\tan(\varphi_{g})=\frac{1}{\cos(\chi_{i})}$ (or $\varphi_{g}=50.3\deg$)
and $\tan(\varphi_{g})=\frac{-1}{\cos(\chi_{i})}$ (or $\varphi_{g}=-50.3\deg$)
respectively. This would ensure that the unwanted transmitted field
from the first grid would almost be entirely transmitted through the
second grid, therefore getting rid of the resonance.

Obviously, trying to solve for such a configuration using Maxwell's
equations would be a formidable task. We can however use our scattering
matrix model developed in section \ref{sec:imp}. We then have to
define two matrices, one for each grid, and solve the problem for
cases where they are separated by a given distance while terminating
the last grid by the line characteristic impedance $Z_{o}$ ($E_{3}^{+}=E_{4}^{+}=0$).
We simulated the response predicted for such an arrangement of grids
and got results that are practically identical to those presented
in Figure \ref{fig:comp} for the optimum polarizer.

\section{Grids and beams of radiation.}

Until now we have restricted our analysis to cases where the dimensions
of the grid (or the assembly) and the extent of the incident wave
were assumed to be infinite. These simplifications were necessary
in order to allow us to have a chance at a solution, as the reality
of finite sizes brings severe difficulties in the analysis. It would
however seem reasonable to suppose that if the incoming excitation
can be properly represented by a beam of radiation which is of a size
a few times smaller that the actual dimensions of the assembly that
the results obtained with our analysis should still be valid. Indeed,
one could argue that the incident beam should only induce currents
in the vicinity of the area where it impinges on the assembly. There
should therefore be little to no differences in its response whether
it is infinite or not.

Although we believe this argument to be a reasonable one, we will
show that the characteristics of the incoming radiation can be important
in some cases. We will in fact argue that it can explain the discrepancies
in the width and shape of the resonance observed in the response of
the reflecting polarizer presented in section \ref{sec:exp} (see
Figure \ref{fig:results}).

As a starting point, let's take note that we can always mathematically
express a beam of radiation $E_{o}(\mathbf{r})$ as a summation of
plane waves with different amplitude (and phase) and $\mathbf{k}$
vectors. For example, in the laboratory system of coordinates ($u$,$v$,$w$),
the electric field along the $u$-axis $E_{i}^{u}(\mathbf{r},t)=\alpha''E_{o}(\mathbf{r})\exp\left(j\omega_{o}t\right)$
can be expressed by (using its Fourier transform in ($\omega,\mathbf{k}$)-space
and assuming the beam to be monochromatic at $\omega_{o}=kc$):

\begin{eqnarray}
E_{i}^{u}(\mathbf{r},t) & = & \frac{\alpha''}{\left(2\pi\right)^{4}}\int_{-\infty}^{\infty}d\omega\, d^{3}kE_{o}(\mathbf{k},\omega)\exp\left[-j\left(\mathbf{k}\cdot\mathbf{r}-\omega t\right)\right]\label{eq:IFT}\\
E_{o}(\mathbf{k},\omega) & = & 2\pi\delta\left(\omega-\omega_{o}\right)\int_{-\infty}^{\infty}d^{3}r'E_{o}(\mathbf{r}')\exp\left(j\mathbf{k}\cdot\mathbf{r}'\right)\label{eq:FT}\end{eqnarray}

\noindent where $\delta\left(x\right)$ is Dirac's delta distribution.

Let's now assume that the incident radiation can be satisfactorily
modeled using a circular Gaussian beam with a beam waist $W_{o}$
and a Rayleigh range $z_{R}=\frac{\pi W_{o}^{2}}{\lambda}$. We also
know that the resonance will occur for a grid-mirror separation of
$h\simeq\frac{\pi}{k_{z}}$ for each spectral component, where $k_{z}$
is the projection of the wave vector along the $z$-axis perpendicular
to the reflecting polarizer (in the coordinates system of the grid
of Figure \ref{fig:wire}). From this we can express the width of
the resonance $\Delta h$ as a function of $k_{z}$ and $\Delta k_{z}$
the spectrum extent along the same axis:\[
\Delta h\simeq\pi\frac{\Delta k_{z}}{k_{z}^{2}}\,.\]

We need to find an expression for $\Delta k_{z}$ and this can be
done as follows. Using the wave uncertainty relation, we can evaluate
the spectrum extent in the laboratory system of axes as:

\begin{eqnarray*}
\Delta k_{u} & \simeq & W_{o}^{-1}\\
\Delta k_{v} & \simeq & W_{o}^{-1}\\
\Delta k_{w} & \simeq & z_{R}^{-1}\,.\end{eqnarray*}

Transforming these in the coordinate system of the grid and inserting
the result in the expression for the width of the resonance we find:

\begin{equation}
\Delta h\simeq\frac{\pi}{\left(k\gamma W_{o}\right)^{2}}\sqrt{W_{o}^{2}\left(1-\gamma^{2}\right)+\left(\frac{2\gamma}{k}\right)^{2}}\,.\label{eq:width}\end{equation}

When we use the corresponding values of the different parameters appearing
in equation (\ref{eq:width}) for the case of the reflecting polarizer
discussed in the last section ($\lambda=1.3$ mm, $\gamma=0.83$ and
$W_{o}=3$ mm) we find $\Delta h\simeq37\,\mu$m. If we take into
account the finite bandwidth of the OVRO receivers (1 GHz), a similar
exercise shows that at most only a few microns need to be added to
the previous estimate. Although these numbers represent only a rough
calculation of what could be expected, they nevertheless tell us that
there will be a significant broadening of the resonant feature. 

We will not try to produce a perfect fit to the data obtained at OVRO
for this would require extensive modeling of our experimental set-up
and therefore bring us to a level of complexity that we do not wish
to tackle at this time. But using equations (\ref{eq:IFT}) and (\ref{eq:FT})
and applying the result of our analysis of the reflecting polarizer
for every spectral component hence calculated, we can get a better
idea of the phenomenon considered here. 

\begin{figure}
\begin{center}\plotone{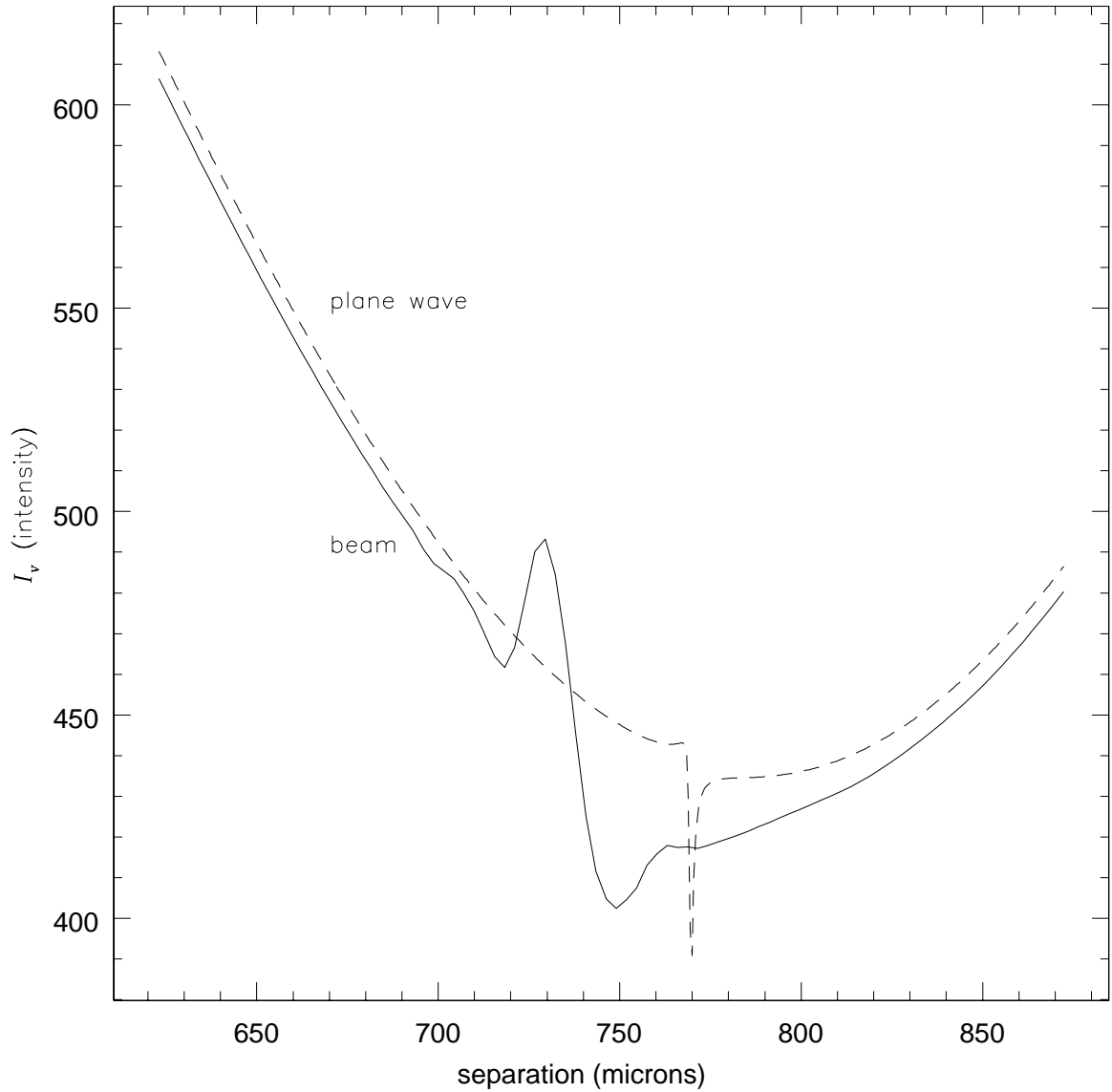}\end{center}

\caption{\label{fig:beam}Simulation of the effect of a Gaussian beam ($W_{o}=3$
mm) on the width and shape of the resonance exhibited by a reflecting
polarizer as discussed in section \ref{sec:exp}. The broken and solid
curves show the results predicted for an incident plane wave and a
Gaussian beam respectively.}
\end{figure}

We have done this and the result is shown in Figure \ref{fig:beam}
where we present the result of a simulation of the effect of a Gaussian
beam on the width and shape of the resonance exhibited by a reflecting
polarizer of the kind discussed in section \ref{sec:exp}. The beam
is converging with its waist situated some 10 cm {}``behind'' the
polarizer and the integrated power (over a beam width) is measured
some distance away from the assembly in the far-field. Although there
still remain some differences, this simulation shares a lot of the
same features observed experimentally. 

We believe that simulations like this one along with our earlier calculations
provide convincing and compelling evidences for the importance of
appropriately taking into account the nature of the incident radiation
in the analysis of similar systems.

\section{Conclusion.}

In this paper, a general solution for the analysis of polarizing grids
was presented; it is valid for arbitrary angles of incidence and of
grid rotation. With it and the scattering matrix representation that
derives from it, basically any configuration or system of grids can
be analyzed as long as some assumptions concerning the wire radius
and spacing are respected ($\lambda>40\, a$ and $d>4\, a$). This
is not a severe restriction as most grid currently available satisfy
those conditions, we refer the reader to \citet{Chambers 1986, Chambers 1988}
for cases where a larger size of wire is needed. Our analysis also
allowed us to define a set of optimum values for both the wire radius
and spacing as specified by the following equations: 

\begin{eqnarray*}
a & \simeq & \left[\frac{\lambda^{5}}{\left(1-\alpha^{2}\right)^{4}\pi^{7}\sigma Z_{o}}\right]^{\frac{1}{6}}\\
d & \simeq & 2\pi a\,.\end{eqnarray*}

We provided an analysis of the effects that two types of random errors
can have on the performance of a grid. It was shown that errors in
the wire spacing were the most important and could have some impact
on the amount of unwanted polarization transmitted through a grid.
In that respect, our model showed to be in good agreement with the
experimental results of \citet{Shapiro}.

Comparisons with experimental data obtained in the calibration of
a reflecting polarizer used at the OVRO were also presented and predictions
from our model are in good agreement with it. The only discrepancies
appeared in the nature of a resonance, more precisely its width. But
we have shown that it could be accounted for by including in the analysis
a proper treatment of the effects of the nature of the incident radiation
on the response of the polarizer.

We are grateful to J. B. Shapiro and E. E. Bloemhof for their permission
to use their previously published experimental results. We wish to
thank the staff of the Owens Valley Radio Observatory, and O. P. Lay
 for numerous discussions and suggestions. The Owens Valley Radio
Observatory is funded by the National Science Foundation under Contract
No. AST 96-13717 and the polarimetry project at OVRO through NASA
grant NAG5-4462. M. H. work was supported in part by a grants from
FCAR and the département de physique de l'Université de Montréal. 

\pagebreak

\appendix

\section{Generalization of the principal axes.}

Referring to equations (\ref{eq:Smatrix}) for the scattering matrix,
one first realizes that such a symmetry in its components will always
be seen when representing an arbitrary polarizing grid. Only the functions
which define $R^{uu}$, $T^{uu}$, \ldots{} will change. Further,
when trying to reduce the scattering matrix to a form similar to equation
(\ref{eq:Sd}) it is only necessary to concentrate on only one of
the two blocks (each appearing twice) present in equation (\ref{eq:Smatrix}).
For example, if one diagonalizes the block composed of the reflection
coefficients then the transmission block is also diagonalized and
vice-versa. In obtaining the results which follow, we have worked
with the reflection block appearing in the upper left and lower right
of equation (\ref{eq:Smatrix}). We will now show that the orientation
of the principal axes is determined by the symmetry of the grid and
can be deduced using the formalism of group theory. 

As seen by the incident wave, the grid has a symmetry which can be
expressed by a representation of the point group $C_{2v}$. The four
covering involved are: the identity ($E$), a rotation by $\pi$ about
the $w$-axis ($C_{2}$), a reflection ($\sigma_{v}$) across a plane
defined by the $w$-axis and and an axis defined by the projection
of the direction of the wires in the plane of the incident field and
finally another reflection ($\sigma_{v}'$) across a plane perpendicular
to the previous one (and to the plane of the incident field). Upon
studying the character table of this group (see \citet[p. 325]{Tinkham})
and the effect of the above operations on the two possible states
of linear polarization (along the $u$ and $v$-axes) we find that
only two non degenerate irreducible representations ($B_{1}$ and
$B_{2}$ in \citet[p. 325]{Tinkham}) will be realized. For each of
these there will exist one eigenvector, each corresponding to a given
principal axis. These can be deduced by constructing the appropriate
symmetry coordinates \citep{Wilson} which turn out to be the two
principal axes $\mathbf{p}_{1}$ and $\mathbf{p}_{2}$ previously
defined in section \ref{sec:axis}. 

Since this result was obtained with the use of group theory, it is
perfectly general and independent of any approximations that can be
used in dealing with a polarizing grid.\pagebreak

\section{List of symbols.}

\begin{figure}
\begin{center}\plotone{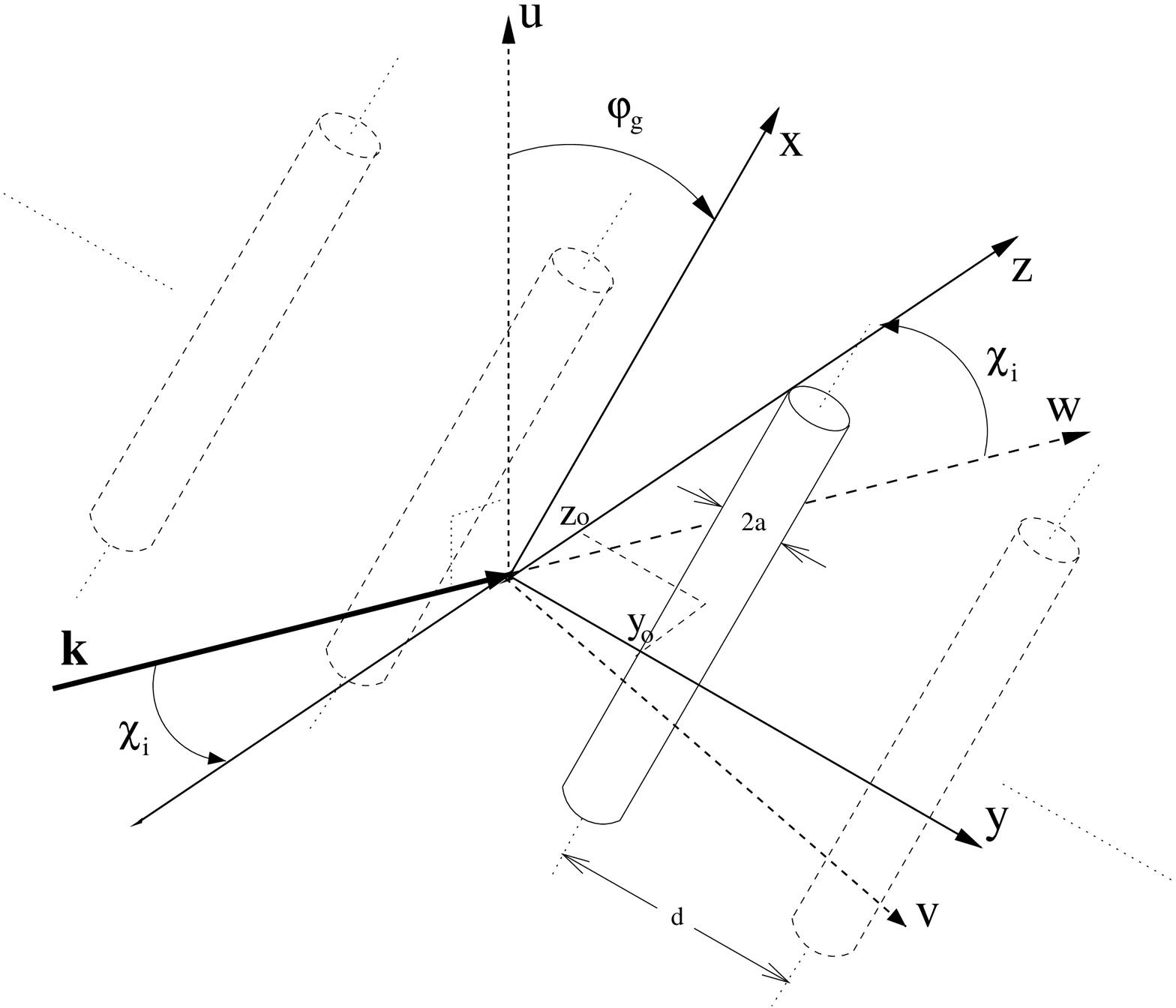}\end{center}

\caption{Coordinates system for the study of a polarizing grid or a single
wire. The wave vector $\mathbf{k}$ of the incident radiation is aligned
with the $w$-axis, the $u$, $x$ and $y$-axes are in the plane
of the page, the $w$ and $z$-axes are in the plane perpendicular
to the $u$-axis (into the page) and the wires are parallel to the
$xy$-plane. We refer to the ($u,v,w$) and ($x,y,z$) systems as
the laboratory and grid coordinates respectively.\label{fig:wire}}
\end{figure}

\begin{figure}
\begin{center}\plotone{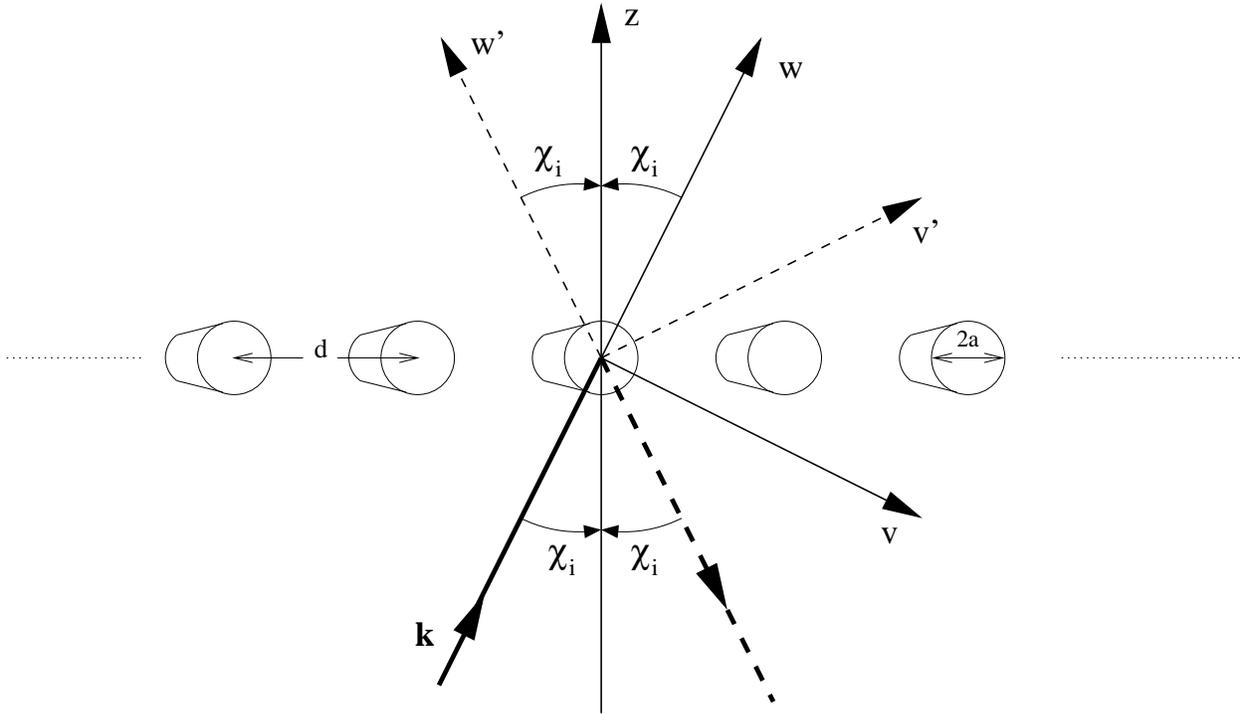}\end{center}

\caption{Definition of the system of coordinates ($u',v',w'$) for the reflected
wave in relation to the ($u,v,w$) system of the incident/transmitted
waves introduced earlier in Figure \ref{fig:wire}. The $u'$ and
$u$-axes are one and the same and are pointing out of the page. The
direction of propagation of the reflected wave is along the negative
$w'$-axis.\label{fig:u'v'w'}}
\end{figure}

\begin{itemize}
\item $a$ wire radius,
\item $\mathbf{A}_{s}$ scattering vector potential,
\item $c$ speed of light in free space,
\item $d$ wire spacing,
\item $\mathbf{E}_{i},\mathbf{E}_{s},\mathbf{E}_{T}$ incident, scattered
and total electric field,
\item $F$ constant $\left(=\frac{\pi\mu_{o}\omega a}{2}\right)$,
\item $\mathbf{F}_{s}$ scattering vector potential,
\item $h$ grid-mirror separation (reflecting polarizer),
\item $\mathbf{H}_{s}$ scattered magnetic field,
\item $H_{n}^{(2)}$ Hankel function of the second kind of order $n$,
\item $\mathbf{J}$ \textbf{}current density vector,
\item $J_{n}$ Bessel function of the first kind of order $n$,
\item $\mathbf{k}$ wave vector of the incident wave $\left(|\mathbf{k}|=k=\frac{2\pi}{\lambda}\right)$,
\item $k'$ $=k\sqrt{1-\alpha^{2}}$,
\item $\mathbf{K}$ total surface current density vector,
\item $\overline{K^{x}},\overline{K^{\theta}}$ mean longitudinal and azimuthal
surface current densities,
\item $\mathbf{p}_{1},\mathbf{p}_{2}$ principal axes of a grid,
\item $R^{x},R^{y},R^{z}$ reflection coefficients in the system of coordinates
of the grid (see Figure \ref{fig:wire}),
\item $R^{u'},R^{v'}$ reflection coefficients in the system of coordinates
of the laboratory (see Figures \ref{fig:wire} and \ref{fig:u'v'w'}),
\item $R_{\Vert},R_{\bot}$ reflection coefficients along the principal
axes of a grid,
\item $R_{TE},R_{TM}$ transverse electric and transverse magnetic reflection
coefficients of the mirror,
\item $\mathbf{S}$ scattering matrix,
\item $T^{x},T^{y},T^{z}$ transmission coefficients in the system of coordinates
of the grid (see Figure \ref{fig:wire}),
\item $T^{u},T^{v}$ transmission coefficients in the system of coordinates
of the laboratory (see Figures \ref{fig:wire} and \ref{fig:u'v'w'}),
\item $T_{\Vert},T_{\bot}$ transmission coefficients along the principal
axes of a grid,
\item $W_{o},z_{R}$ beam waist and Rayleigh range of a circular Gaussian
beam,
\item $\mathbf{Z}$ impedance matrix,
\item $Z_{o},$ impedance of free space $\left(=\sqrt{\frac{\mu_{o}}{\varepsilon_{o}}}\right)$,
\item $Z_{p},Z_{n}$ grid impedance along the principal axes (as de\-fined
with the im\-pe\-dance matrix), 
\item $Z_{s}$ surface impedance of the wires $\left(=\left(1+j\right)\sqrt{\frac{\mu_{o}\omega}{2\sigma}}\right)$,
\item $Z_{\Vert},Z_{\bot}$ grid impedance along the principal axes (as
defined with the scattering matrix), 
\item $\alpha$ projection of the nor\-ma\-lized wa\-ve vec\-tor on
the $x$-axis \\
$\left(=\sin\left(\chi_{i}\right)\sin\left(\varphi_{g}\right)\right)$,
\item $\beta$ projection of the nor\-ma\-lized wa\-ve vec\-tor on the
$y$-axis \\
$\left(=\sin\left(\chi_{i}\right)\cos\left(\varphi_{g}\right)\right)$,
\item $\gamma$ projection of the normalized wave vector on the $z$-axis
$\left(=\cos\left(\chi_{i}\right)\right)$,
\item $\alpha',\beta',\gamma'$ projection of the normalized incident field
on the $x$, $y$ and $z$-axes,
\item $\alpha'',\beta''$ projection of the normalized incident field on
the $u$ and $v$-axes,
\item $\delta\left(x\right)$ Dirac's delta distribution,
\item $\varepsilon_{o},\mu_{o}$ permittivity and permeability of free space,
\item $\eta,\xi$ random errors in wire radius and spacing,
\item $\lambda$ wavelength,
\item $\sigma$ wire conductivity,
\item $\varphi_{g},\chi_{i}$ angle of grid rotation and angle of incidence,
\item $\Psi$ Euler's constant $\left(\simeq0.577215\right)$,
\item $\omega$ angular frequency of radiation.
\end{itemize}
\pagebreak


\end{document}